\documentclass[reqno,a4paper]{amsart}
\usepackage[active]{srcltx}
\usepackage{times}
\usepackage{mathrsfs}
\usepackage{latexsym,amsopn,amssymb,amsmath}
\usepackage{amsthm}
\usepackage{amsfonts,amsbsy,amscd,stmaryrd}
\usepackage{paralist}
\usepackage{geometry}
\oddsidemargin %-0.1cm
\evensidemargin %-0.1cm
\newcommand{\C}{\mathbb{C}}

\newcommand{\mbn}{\mathbb{N}}
\newcommand{\R}{\mathbb{R}}

\newcommand{\mbz}{\mathbb{Z}}
\newcommand{\ca}{\mathcal{A}}

\newcommand{\cc}{\mathcal{C}}

\newcommand{\ce}{\mathcal{E}}
\newcommand{\cf}{\mathcal{F}}

\newcommand{\ch}{\mathcal{H}}
\newcommand{\ci}{\mathcal{I}}

\newcommand{\co}{\mathcal{O}}
\newcommand{\cp}{\mathcal{P}}

\newcommand{\cu}{\mathcal{U}}

\def\rond{\mathscr}
\newcommand{\ra}{\rond{A}}
\newcommand{\rb}{\rond{B}}
\newcommand{\rc}{\rond{C}}

\newcommand{\re}{\rond{E}}

\newcommand{\rg}{\rond{G}}

\newcommand{\rj}{\rond{J}}
\newcommand{\rk}{\rond{K}}
\newcommand{\rl}{\rond{L}}
\newcommand{\rM}{\rond{M}}

\newcommand{\rn}{\rond{N}}

\def\rmb{\mathrm{b}}
\def\rmc{\mathrm{c}}
\def\c{\mathrm{c}}
\def\coa{\mathrm{co}}

\def\d{\text{d}}
\def\rme{\mathrm{e}}
\def\rmi{\text{i}}
\def\rmo{\mathrm{o}}
\def\rmr{\mathrm{r}}
\newcommand{\Cb}{\cc_{\rmb}}

\newcommand{\ccfr}{\cc_{\mathrm{trl}}}
\newcommand{\Co}{\cc_{\rmo}}
\newcommand{\Cc}{\cc_{\rmc}}

\def\vphi{\varphi}
\def\vkappa{\varkappa}
\newcommand{\ind}{1} %{\mbox{\bf 1}}
\renewcommand{\proof}{\noindent{\bf Proof: }}

\def\bra#1{\langle{#1}|}
\def\ket#1{|{#1}\rangle}
\def\braket#1#2{\langle{#1}|{#2}\rangle}

\def\spe{\mathrm{Sp_{ess}}}
\def\sp{\mathrm{Sp}}

\def\what{\widehat}
\def\what#1{\widehat{ #1\,}}
\def\wtilde{\widetilde}

\def\nin{\notin}

\def\supp{\mbox{\rm supp\! }}
\def\ess{\mathrm{ess}}
\def\spec{\mathrm{Sp}}
\def\nin{\notin}
\def\pprod{\textstyle \prod}

\def\ccup{\textstyle{\bigcup}}
\def\ccap{\textstyle{\bigcap}}

\def\qed{\hfill \raisebox{0.5ex}{\framebox[1.6ex]{
                                       \rule[0ex]{0ex}{0.3ex} }}}

\def\build#1_#2^#3{\mathrel{\mathop{\kern 0pt#1}\limits_{#2}^{#3}}}
\newcounter{PAR}[section]

\newtheorem{theorem}{Theorem}[section]
\newtheorem{lemma}[theorem]{Lemma}
\newtheorem{proposition}[theorem]{Proposition}

\newtheorem{definition}[theorem]{Definition}
\newtheorem{remark}[theorem]{Remark}

\newtheorem{example}[theorem]{Example}

\newtheorem*{acknowledgment}{Acknowledgment}
\long\def\symbolfootnote[#1]#2{\begingroup%
\def\thefootnote{\fnsymbol{footnote}}\footnote[#1]{#2}\endgroup} 
\begin{document}

\title[Elliptic operators on metric spaces]{On the structure of the 
essential spectrum \\ of elliptic operators on metric spaces}

\author[Vladimir GEORGESCU]{Vladimir GEORGESCU} 
\address{CNRS and  
University of Cergy-Pontoise 95000 Cergy-Pontoise, France}
\email{vlad@math.cnrs.fr} 

%\date{\today} 

\thanks{}


\begin{abstract}
  We give a description of the essential spectrum of a large class
  of operators on metric measure spaces in terms of their
  localizations at infinity.  These operators are analogues of the
  elliptic operators on Euclidean spaces and our main result
  concerns the ideal structure of the $C^*$-algebra generated by
  them.
\end{abstract}

\maketitle 

\section{Introduction} \label{s:intro}

\subsection{}

The question we consider in this paper is whether the essential
spectrum of an operator can be described in terms of its
``localizations at infinity''. Later on we give a precise
mathematical meaning to this notion along the following lines: we
first define a $C^*$-algebra $\re$ which should be thought as the
minimal \mbox{$C^*$-algebra} which contains the resolvents of the
operators we want to study, then we point out a remarkable class of
geometrically defined ideals $\re_{(\vkappa)}$ in $\re$, where
$\vkappa$ are certain ultrafilters on $X$, and finally we define the
localization of an operator in $\re$ at $\vkappa$ as its image in
the quotient $C^*$-algebra $\re_\vkappa=\re/\re_{(\vkappa)}$.  For
the moment we shall stick to the naive interpretation of
localizations at infinity of an operator $H$ as ``asymptotic
operators'' obtained as limits of translates of $H$ to infinity, but
we stress that translations have no meaning for the class of spaces
of interest here and very soon we shall abandon this point of view.

We begin with the case $X=\R^d$. Note that we are interested only in
operators $H$ which are self-adjoint (Hamiltonians of quantum
systems). Denote $U_a$ the unitary operator of translation by $a\in
X$ in $L^2(X)$, so that $(U_af)(x)=f(x+a)$. We say that $H_\vkappa$
is an asymptotic Hamiltonian of $H$ if there is a sequence $a_n\in
X$ with $|a_n|\to\infty$ such that $U_{a_n}HU_{a_n}^*$ converges in
strong resolvent sense to $H_\vkappa$. Then we have
$\spe(H)=\overline\cup_\vkappa\sp(H_\vkappa)$ for very large classes
of Schr\"odinger operators. We refer to the paper \cite{HM} of
Helffer and Mohamed as one of the first dealing with this question
in a general setting and to that of Last and Simon \cite{LS} for the
most recent results obtained by similar techniques (geometric
methods involving partitions of unity) and for a complete list of
references. We mention that the importance of the asymptotic
operators has been emphasized in a series of papers in the nineties
by Rabinovich, Roch, and Silbermann and summarized in their book
\cite{RRS} (see also \cite{CL}; we thank B.\ Simon for this
reference). They are especially concerned with the case $X=\mbz^d$
and treat differential operators on $L^p(\R^d)$ with the help of a
discretization method.

Results of this nature have also been obtained in \cite{GI0,GI} by a
quite different method where the description of localizations at
infinity in terms of asymptotic operators is not so natural and
rather looks like an accident.  To explain this point, we recall one
result. Let $X$ be an abelian locally compact non-compact group,
define $U_a$ as above, and for any character $k$ of $X$ let $V_k$ be
the operator of multiplication by $k$ on $L^2(X)$. Let
$\re\equiv\re(X)$ be the set of bounded operators $T$ on $L^2(X)$
such that $\|V_k^*TV_k-T\|\to0$ and $\|(U_a-1)T^{(*)}\|\to0$ when
$k\to1$ and $a\to0$.  A self-adjoint operator $H$ satisfying
$(H-i)^{-1}\in\re$ is said to be affiliated to $\re$; it is easy to
see that this class of operators is very large.  Let
$\delta\equiv\delta(X)$ be the set of ultrafilters on $X$ finer than
the Fr\'echet filter. If $H$ is affiliated to $\re$ then for each
$\vkappa\in\delta$ the limit $\lim_{a\to\vkappa}
U_aHU_a^*=H_\vkappa$ exists in the strong resolvent sense and we
have $\spe(H)=\overline\cup_{\vkappa\in\delta}\sp(H_\vkappa)$.  Thus
the essential spectrum of an operator affiliated to $\re$ is
determined by its asymptotic operators.

The proof goes as follows. The space $\re$ is in fact a
$C^*$-algebra canonically associated to $X$, namely the crossed
product $\cc(X)\rtimes X$ of the algebra $\cc(X)$ of bounded
uniformly continuous functions on $X$ by the natural action of
$X$. Moreover, the space $\rk\equiv\rk(X)$ of compact operators on
$L^2(X)$ is an ideal of $\re$.  Note that by ideal in a
$C^*$-algebra we mean ``closed bilateral ideal'' and we call
morphism a $*$-homomorphism between two $*$-algebras. It is easy to
see that for each $\vkappa\in\delta$ and each $T\in\re$ the strong
limit $\tau_\vkappa(T):=\lim_{a\to\vkappa}U_aTU_a^*$ exists and that
the so defined $\tau_\vkappa$ is an endomorphism of $\re$ so its
kernel $\ker\tau_\vkappa$ is an ideal of $\re$ which clearly
contains $\rk$. The main fact is
$\ccap_{\vkappa\in\delta}\ker\tau_\vkappa=\rk$ and and this is the
only nontrivial part of the proof. From here we immediately deduce
the preceding formula for the essential spectrum of the operators
affiliated to $\re$. Indeed, it suffices to recall that the
essential spectrum of an operator in a $C^*$-algebra like $\re$
which contains $\rk$ is equal to the spectrum of the image of the
operator in the quotient algebra $\re/\rk$.

We shall call $\re$ the \emph{elliptic $C^*$-algebra of the group
  $X$}. It is probably not clear that this has something to do with
the elliptic operators, but the following fact justifies the
terminology.  The $C^*$-algebra generated by a set of self-adjoint
operators on a given Hilbert space is by definition the smallest
$C^*$-algebra which contains the resolvents of these operators.  Let
$X=\R^d$ and let $h$ be a real elliptic polynomial of order $m$ on
$X$. Then $\re$ is the $C^*$-algebra generated by the self-adjoint
operators of the form $h(\rmi\nabla)+S$ where $S$ runs over the set
of symmetric differential operators of order $<m$ whose coefficients
are $C^\infty$ functions which are bounded together with all their
derivatives.  We stress that although $\re(X)$ is generated by a
small class of elliptic differential operators, the class of
self-adjoint operators affiliated to it is quite large and contains
many singular perturbations of the usual elliptic operators. This is
obvious from the description of $\re$ we gave before and many
explicit examples may be found in \cite{DG,GI}.

\subsection{}

Our purpose is to extend the framework and the results stated above
to the case when $X$ is a metric space without any group structure
or group action and for which the notion of differential operator is
not a priori defined. To each measure metric space $X=(X,d,\mu)$
satisfying some quite general conditions we associate a
$C^*$-algebra $\re\equiv\re(X)$ of operators on $L^2\equiv
L^2(X,\mu)$ and to each $\vkappa\in\delta(X)$ we associate an ideal
$\re_{(\vkappa)}$ of $\re$ such that
$\ccap_{\vkappa}\re_{(\vkappa)}$ is the space $\rk$ of compact
operators on $L^2$ if the metric space $X$ has a certain amenability
property, namely the Property A of Guoliang Yu \cite{Yu}.  The
$\re_{(\vkappa)}$ are analogues of the $\ker\tau_\vkappa$ and the
image of an operator $T\in\re$ in the quotient algebra
$\re/\re_{(\vkappa)}$ is the analogue of $\tau_\vkappa(T)$. The
ideal $\re_{(\vkappa)}$ is defined in terms of the behavior of the
operators at a region at infinity which contains $\vkappa$.

Our interest in this question was roused by a recent paper of
E.\,B.\ Davies \cite{D} in which a \mbox{$C^*$-algebra} $\rc(X)$,
called \emph{standard algebra}, is associated to each metric measure
space $X$ as above.  Davies points out a class of ideals of $\rc$
and describes their role in understanding the essential spectrum of
the operators affiliated to $\rc$. This algebra is much larger than
$\re$ if $X$ is not discrete. If $X$ is an abelian group as above,
then $\rc$ is the set of bounded operators $T$ on $L^2$ such that
$\|V_k^*TV_k-T\|\to0$ when $k\to1$. It is clearly impossible to give
a complete description of the essential spectrum of such operators
only in terms of their behavior at infinity in the configuration
space $X$ (consider for example the case $X=\R$). A more precise
description of $\rc$ and of its relation with $\re$ may be found in
Section \ref{s:psd}.

In Section \ref{s:groups} we show that if $X$ is a unimodular
amenable group then we have $\re(X)=\cc(X)\rtimes X$ as in the
abelian case.  Thus we may recover as a corollary of our main result
(Theorem \ref{th:detailed}) the results in \cite{GI0,GI} for locally
compact abelian groups and those of Roe \cite{Ro} in the case of
finitely generated discrete (non-abelian) groups (see also
\cite{RRR}). Amenability is not really necessary: in fact, the
natural objects here are the reduced crossed products and then Yu's
Property A is sufficient.

\subsection{}

From a more general point of view, the main point of the approach
sketched above is to shift attention from one operator to an algebra
of operators. Instead of studying the essential spectrum (or other
qualitative spectral properties, like the Mourre estimate) of a
self-adjoint operator $H$ on a Hilbert space $\ch$, we consider a
$C^*$-algebra $\re$ of operators on $\ch$ which contains
$\rk=K(\ch)$ and such that $H$ is affiliated to it and try to find
an ``efficient'' description of the quotient $C^*$-algebra
$\re/\rk$. For this, we look for a family of ideals $\rj_\vkappa$ of
$\re$ such that $\bigcap_\vkappa\rj_\vkappa=\rk$ because then we
have a natural embedding
\begin{equation}\label{eq:eff}
\re/\rk \hookrightarrow \pprod_\vkappa \re/\rj_\vkappa
\end{equation}
and, in our concrete situation, we think of this as an efficient
representation of $\re/\rk$ if the ideals $\rj_\vkappa$ are in some
sense maximal and have a geometrically simple interpretation. This
is in an important point and we shall get back to it later on. For
the moment note that any representation like \eqref{eq:eff} has
useful consequences in the spectral theory of the operators
$T\in\re$, for example if $T$ is normal and $T_\vkappa$ is the
projection of $T$ in $\re/\rj_\vkappa$ then its essential spectrum
is given by
\begin{equation}\label{eq:spe}
\spe(T)=\overline\ccup_\vkappa \sp(T_\vkappa).
\end{equation}
Arbitrary ideals $\rj\subset\re$ also play a role in the spectral
analysis of the operators $T\in\re$. For example, if we denote
$T/\rj$ the image of $T$ in the quotient algebra $\re/\rj$ then
clearly $\sp(T/\rj)\subset\sp(T)$ and if $\rj$ contains the compacts
then $\sp(T/\rj)\subset \spe(T)$. It is natural in our framework to
call the quotient operator $T/\rj$ \emph{localization of $T$ at
  $\rj$} (see Section \ref{s:env} for the meaning of this operation
in the abelian case). Observe that $\sp(T/\rj)$ becomes smaller when
$\rj$ increases, which allows a better understanding of parts of the
spectrum of $T$. In particular, it will become clear later on that
by taking large $\rj$ one can isolate the contribution to the
essential spectrum of $T$ of the localization of $T$ to small
regions at infinity.

We refer to \cite{ABG,BoG1,BoG2,DG2,Ge} for a general discussion
concerning the operation of localization with respect to an ideal
and for applications in the spectral theory of many-body systems and
quantum field theory but we shall mention here an example which is
relevant also in the present context. Let $H$ be the Hamiltonian of
a system of $N$ non-relativistic particles interacting through
two-body potentials and let $V_{jk}$ be the potential linking
particles $j$ and $k$. For each partition $\sigma$ of the system of
particles let $H_\sigma$ be the Hamiltonian obtained by replacing by
zero the $V_{jk}$ such that $j,k$ belong to different clusters of
$\sigma$. Then the HVZ theorem says that $\spe(H)=\ccup_\sigma
\sp(H_\sigma)$ where $\sigma$ runs over the set of two-cluster
partitions.  In fact, this is an immediate consequence of the
preceding algebraic formalism: the $N$-body $C^*$-algebra is easy to
describe and $H_\sigma$ is the localization of $H$ at a certain
ideal which appears very naturally in this context. The point is
that we do not have to take some limit at infinity to get
$H_\sigma$, although this could be done (this would mean that we use
``geometric methods''). The ideals which are involved in the
representation \eqref{eq:eff} in this case are \emph{minimal} in a
precise sense. In particular, the preceding decomposition of the
spectrum is very rough (you do not see the contribution of
$k$-cluster partitions with $k>2$).

In connection with the algebraic approach sketched above, we would
like to emphasize the previous work of J.\ Bellissard, who was one
of the first to stress the advantage of considering $C^*$-algebras
generated by Hamiltonians in the context of solid state physics
\cite{Be1,Be2}, and that of H.\,O.\ Cordes, who studied
$C^*$-algebras of pseudo-differential operators on manifolds and
their quotients with respect to the ideal of compact operators
\cite{Cor} already in the seventies.

\subsection{}

Now let's get back to our problem. Assuming we have chosen the
``correct'' algebra $\re(X)$, we must find the relevant ideals. In
the group case, this is easy, because there is a natural class of
ideals associated to translation invariant filters
\cite{GI0}. Proposition \ref{pr:gcoarse} gives a characterization of
these filters which involves only the metric structure of $X$ (in
fact, only the coarse structure associated to it \cite{R}).  Thus
what we call \emph{coarse filters} in a metric space are analogs of
the invariant filters in a group. To each coarse filter $\xi$ we
then associate an ideal $\rj_\xi$ defined in terms of the behavior
of the operators at a certain region at infinity defined by $\xi$,
cf.\ \eqref{eq:ek1*}. These are the geometric ideals which play the
main role in or analysis.

% The next comments should clarify this point.

Recall that the set of ultrafilters finer than the Fr\'echet filter
is a compact subset $\delta(X)$ of the Stone-\v{C}ech
compactification $\beta(X)$ of $X$. Any filter $\xi$ finer than
Fr\'echet can be thought as a closed subset of $\delta(X)$ by
identifying it with the set $\xi^\dagger$ of ultrafilters finer than
it, and then the sets $F\in\xi$ can be thought as traces on $X$ of
neighborhoods of this closed set in $\beta(X)$. The sets
$\xi^\dagger$ with $\xi$ coarse will be called \emph{coarse subsets}
of $\delta(X)$ (they are closed).  If $X$ is a group then $X$ acts
on $\delta(X)$, the coarse subsets are the closed invariant subsets
of $\delta(X)$, and the small invariant sets are parametrized as
follows: to each $\vkappa\in\delta(X)$ we associate the smallest
closed invariant set containing $\vkappa$ (i.e.\ the closure of the
orbit which passes through it). But this can be easily expressed in
group independent terms: if $\vkappa\in\delta(X)$ let
$\coa(\vkappa)$ be the finer coarse filter included in $\vkappa$ and
let $\what\vkappa:=\coa(\vkappa)^\dagger$ be the smallest coarse set
containing $\vkappa$. Then the $\coa(\vkappa)$ are the large coarse
filters, the $\what\vkappa$ the small coarse sets, and the
$\re_{(\vkappa)}:=\rj_{\coa(\vkappa)}$ are the large coarse ideals
which should allow us to compute the essential spectrum of the
operators in $\re$. Heuristically speaking, $\re_{(\vkappa)}$
consists of the operators in $\re$ which vanish at $\what\vkappa$.
For example, if $X$ is discrete, so $\re$ contains the bounded
functions $\varphi$ on $X$, we have $\varphi\in\re_{(\vkappa)}$ if
and only if the continuous extension of $\varphi$ to $\beta(X)$ is
zero on $\what\vkappa$.

We stress that this strategy denotes a certain bias toward the role
played by the behavior at infinity in $X$ (thought as physical or
configuration space): we think that it has a dominant role since we
hope that our choices of ideals is sufficient to describe the
quotient $\re/\rk$. There is no a priori reason for this to be true:
there are physically natural situations in which ideals defined in
terms of behavior at infinity in momentum or phase space must be
taken into account \cite{GI0}. However, it does not seem so clear to
us how to define such physically meaningful objects in the present
context (there is no natural phase space). 

Anyway, the situation is not simple even at the level of
geometrically defined ideals. Indeed, the ideals $\re_{(\vkappa)}$
are defined in terms of the behavior of the operators in $\re$ at
$\what\vkappa$, but it is not completely clear how to express the
intuitive idea that an operator $T$ vanishes on $\what\vkappa$. Our
choice is the more restrictive one, but there is a second one which
is also quite natural and leads to a distinct class of ideals
$\rg_\vkappa$, cf.\ \eqref{eq:ek1} and \eqref{eq:gxi}.  One has
$\re_{(\vkappa)}\subset\rg_\vkappa$ strictly in general but equality
holds if the space $X$ has the Property A.

In general the ideals $\rg_\vkappa$ do not suffice to compute $\rk$,
i.e. \emph{we do not have}
$\ccap_{\vkappa\in\delta}\rg_{\vkappa}=\rk$.  In fact an ideal $\rg$
which contains the compacts appears naturally in the algebra $\re$,
the so-called \emph{ghost ideal}, and this ideal could contain a
projection of infinite rank, hence be strictly larger than the
compacts. The construction of such a projection is due to Higson,
Laforgue and Skandalis \cite{HLS} and is important in the context of
the Baum-Connes conjecture.  They consider the simplest case of
discrete metric spaces with bounded geometry (the number of points
in a ball of radius $r$ is bounded independently of the center of
the ball) when $\re$ is the \emph{uniform Roe $C^*$-algebra}. More
information concerning this question may be found in the papers
\cite{CW1,CW2,Wa} by Chen and Wang where the ideal structure of the
uniform Roe algebra \cite{R} is studied in detail.  Their idea of
using kernel truncations with the help of positive type functions in
case $X$ has Yu's Property A plays an important role in our proofs,
as we shall see in Section \ref{s:ell}.  But before going into
details on these matters we shall describe in the next section in
precise terms the framework and the main results of this paper.  

As explained before, a representation like \eqref{eq:eff} involving
ideals which are as large as possible will provide the most detailed
information on the structure of the essential spectrum of the
observables affiliated to $\re$. Thus the fact that
$\ccap_{\vkappa\in\delta}\rg_{\vkappa}\neq\rk$ shows that in general
the large ideals are not sufficient to compute the essential
spectrum.  We leave open the question whether
$\ccap_{\vkappa\in\delta}\re_{(\vkappa)}=\rk$ holds even if
$\ccap_{\vkappa\in\delta}\rg_{\vkappa}\neq\rk$.

\section{Main results} \label{s:main}

A metric space $X=(X,d)$ is \emph{proper} if each closed ball
$B_x(r) =\{y \mid d(x,y)\leq r\}$ is a compact set. This implies the
local compactness of the topological space $X$ but is much more
because local compactness means only that the small balls are
compact. In particular, if $X$ is not compact, then the metric
cannot be bounded.  We are interested in proper non-compact metric
spaces equipped with Radon measures $\mu$ with support equal to $X$,
so $\mu(B_x(r))>0$ for all $x\in X$ and all $r>0$, and which satisfy
(at least) the following condition
\begin{equation}\label{eq:xrv}
  V(r):=\sup_{x\in X}\mu(B_x(r)) < \infty  \text{ for all real } r>0.
\end{equation}
We shall always assume that a metric measure space $(X,d,\mu)$
satisfies these conditions. On the other hand, for the proof of our
main results we need the following supplementary condition:
\begin{equation}\label{eq:12}
\inf_x\mu(B_x(1/2))>0.
\end{equation} 
The choice of $1/2$ in (i) is, of course, rather arbitrary, and an
assumption of the form $\inf_x\mu(B_x(r))>0$ for all $r>0$ would be
more natural. Each time we use \eqref{eq:12} we shall mention it
explicitly.

To simplify the notations we set $\d\mu(x)=\d x$,
$L^2(X)=L^2(X,\mu)$, and $B_x=B_x(1)$.  We denote $\rb(X)$ the
$C^*$-algebra of all bounded operators on $L^2(X)$ and $\rk(X)$ the
ideal of $\rb(X)$ consisting of compact operators.  For $A\subset X$
we denote $\ind_A$ its characteristic function and if $A$ is
measurable then we use the same notation for the operator of
multiplication by $\ind_A$ in $L^2(X)$.

Several versions of Yu's Property A appear in the literature (see
\cite[Definition 11.35]{R} and \cite{Tu} for the discrete case), we
have chosen that which was easier to state and use in our
context. Later on we shall state and use a more abstract version
which can easily be reformulated in terms of positive type functions
on $X^2$. See page \pageref{p:amen} here and \cite[Ch.\ 3]{R} for the relation with
amenability in the group case.

\begin{definition}\label{df:yu}
  We say that the metric measure space $(X,d,\mu)$ \emph{has
    Property A} if for each $\varepsilon,r>0$ there is a Borel map
  $\phi:X\to L^2(X)$ with $\|\phi(x)\|=1$, $\supp \phi(x)\subset
  B_x(s)$ for some number $s$ independent of $x$, and such that
  $\|\phi(x)-\phi(y)\|<\varepsilon$ if $d(x,y)<r$.
\end{definition}

\begin{definition}\label{df:classa}
  We say that $X=(X,d,\mu)$ is \emph{a class A space} if $(X,d)$ is
  a proper non-compact metric space and $\mu$ is a Borel measure on
  $X$ such that: (i) $\mu(B_x(r))>0$ and $\sup_x\mu(B_x(r))<\infty$
  for each $r>0$, (ii) $\inf_x\mu(B_x(1/2))>0$, (iii) $(X,d,\mu)$
  has property A.
\end{definition}

Since $X$ is locally compact the spaces $\Co(X)$ and $\Cc(X)$ of
continuous functions on $X$ which tend to zero at infinity or have
compact support respectively are well defined.  We use the slightly
unusual notation $\cc(X)$ for the set of \emph{bounded uniformly
  continuous} functions on $X$ equipped with the sup norm. Then
$\cc(X)$ is a $C^*$-algebra and $\Co(X)$ is an ideal in it.  We
embed $\cc(X)\subset\rb(X)$ by identifying $\varphi\in\cc$ with the
operator $\varphi(Q)$ of multiplication by $\varphi$ (this is an
embedding because the support of $\mu$ is equal to $X$). We shall
however use the notation $\varphi(Q)$ if we think that this is
necessary for the clarity of the text.

Functions $k:X^2\to\C$ on the product space $X^2=X\times X$ are also
called kernels on $X$. We say that $k$ is a \emph{controlled kernel}
if there is a real number $r$ such that $d(x,y)>r \Rightarrow
k(x,y)=0$. With the terminology of \cite{HPR}, a kernel is
controlled if it is supported by an entourage of the bounded coarse
structure on $X$ coming from the metric.  We denote $\ccfr(X^2)$ the
set of \emph{bounded uniformly continuous controlled kernels} and to
each $k\in\ccfr(X^2)$ we associate an operator $Op(k)$ on $L^2(X)$ by
$(Op(k)f)(x)=\int_X k(x,y) f(y) \d y$. It is easy to check (see
Section \ref{s:ell}) that the set of such operators is a
$*$-subalgebra of $\rb(X)$. Hence
\begin{equation}\label{eq:ell}
\re(X) \equiv \re(X,d,\mu)= \text{ norm closure of } 
\{Op(k) \mid k\in\ccfr(X^2) \}
\end{equation}
is a $C^*$-algebra of operators on $L^2(X)$. We shall say that
$\re(X)$ is the \emph{elliptic algebra} of $X$.

\begin{remark}\label{re:form}{\rm The following alternative presentation of the
    framework clarifies the role of the metric.  Fix a couple
    $X=(X,\mu)$ consisting of a locally compact non-compact
    topological space $X$ equipped with a Radon measure $\mu$ with
    support equal to $X$. This fixes the Hilbert space
    $L^2(X)$. Then to each proper metric compatible with the
    topology of $X$ and such that $\sup_x\mu(B_x(r))<\infty$ for all
    $r$ we associate a \mbox{$C^*$-algebra} $\re(X,d)$ of operators
    on $L^2(X)$ which contains $\rk(X)$. It is interesting to note
    that $\re(X,d)$ \emph{depends only on the coarse equivalence
      class of the metric}. Recall that two metrics $d,d'$ are
    coarse equivalent if there are positive increasing functions
    $u,v$ such that $d\leq u(d')$ and $d'\leq v(d)$. This can also
    be expressed in terms of coarse structures on $X$ \cite[page
    810]{STY}.  }\end{remark}

There is an obvious $\cc(X)$-bimodule structure on $\re(X)$ and we have
$$\rk(X)=\Co(X)\re(X)=\re(X)\Co(X)\subset\re(X).
$$
As explained in the introduction we are interested in a
``geometrically meaningful'' representation of the quotient
\mbox{$C^*$-algebra} $\re(X)/\rk(X)$. For this we introduce the
class of ``coarse ideals'' described below.

If $F\subset X$ and $r>0$ is real we denote $F^{(r)}$ the set of
points $x$ which belong to the interior of $F$ and are at distance
larger than $r$ from the boundary, more precisely $\inf_{y\nin
  F}d(x,y)>r$. A filter $\xi$ of subsets of $X$ will be called
\emph{coarse} if $F\in\xi\Rightarrow F^{(r)}\in\xi$ for all
$r$. Note that the set of complements of a coarse filter is a coarse
ideal of subsets of $X$ in the sens of \cite{HPR}.  The
\emph{Fr\'echet filter}, i.e.\ the set of sets with relatively
compact complement, is clearly coarse, we denote it $\infty$.  There
is a trivial coarse filter, namely $\xi=\{X\}$, which is of no
interest for us.  All the other coarse filters are finer that
$\infty$.

To each coarse filter $\xi$ on $X$ we associate an ideal of $\re(X)$
by defining
\begin{equation}\label{eq:ek1*}
\rj_\xi(X)=
\{T\in\re(X) \mid \inf_{F\in\xi}\|\ind_F T\|=0\}=
\{T\in\re(X) \mid \inf_{F\in\xi}\|T\ind_F\|=0\}
\end{equation} 
where the $\inf$ is taken only over measurable $F\in\xi$.  We shall
see that the set $\ci_\xi(X)$ of $\varphi\in\cc(X)$ such that
$\lim_\xi \varphi=0$ is an ideal of $\cc(X)$ and
$\rj_\xi(X)=\ci_\xi(X)\re(X)= \re(X)\ci_\xi(X)$.

Let $\beta(X)$ be the set of all ultrafilters of $X$ (this is the
Stone-C\v{e}ch compactification of the \emph{discrete} space $X$)
and let $\delta(X)$ be the set of ultrafilters finer than the
Fr\'echet filter. For each $\vkappa\in\beta(X)$ we denote
$\coa(\vkappa)$ the largest coarse filter contained in $\vkappa$ and
we set $\cc_{(\vkappa)}(X)=\ci_{\coa(\vkappa)}(X)$ and
$\re_{(\vkappa)}(X)=\rj_{\coa(\vkappa)}(X)$.  These are ideals in
$\cc(X)$ and $\re(X)$ respectively and we have
\begin{equation}\label{eq:ideals11}
\re_{(\vkappa)}(X)=\cc_{(\vkappa)}(X)\re(X)= \re(X)\cc_{(\vkappa)}(X).
\end{equation}
If $X$ is of class A then from Theorem \ref{th:nyx} we get a second
description of these ideals.

\begin{proposition}\label{pr:evk}
  If $X$ is a space of class A then for any $\vkappa\in\delta(X)$ we
  have
\begin{equation}\label{eq:ek}
\re_{(\vkappa)}(X)=
\{T\in\re(X) \mid \lim_{x\to\vkappa}\|\ind_{B_x(r)} T\|=0\ \forall
r>0 \}.
\end{equation}
\end{proposition}

Then to each ultrafilter $\vkappa\in\delta(X)$ we associate the
quotient $C^*$-algebra
\begin{equation}\label{eq:local}
\re_\vkappa(X)=\re(X)/\re_{(\vkappa)}(X)
\end{equation} 
and call it \emph{localization of $\re(X)$ at $\vkappa$} We denote
$\vkappa.T$ the image of $T\in\re(X)$ through the canonical morphism
$\re(X)\to\re_{\vkappa}(X)$ and we say that $\vkappa.T$ is the
\emph{localization of $T$ at $\vkappa$}. Our main result is:

\begin{theorem}\label{th:detailed}
  If $X$ is a class A space then $\ccap_{\vkappa\in\delta(X)}
  \re_{(\vkappa)}(X) = \rk(X)$, hence
\begin{equation}\label{eq:detail}
\re(X)/\rk(X)\hookrightarrow \pprod_{\vkappa\in\delta(X)}\re_\vkappa.
\end{equation}
In particular, the essential spectrum of any normal operator
$T\in\re(X)$ is equal to the closure of the union of the spectra of
its localizations at infinity:
\begin{equation}\label{eq:ess}
\spec_\ess(T)=\overline{\ccup}_{\vkappa\in\delta(X)}\spec(\vkappa.T).
\end{equation}
\end{theorem}

In view of applications to self-adjoint operators affiliated to
$\re(X)$, we recall \cite{ABG} that an \emph{observable affiliated
  to a $C^*$-algebra} $\ra$ is a morphism $H:\Co(\R)\to\ra$. We set
$\varphi(H):=H(\varphi)$. If $\cp:\ra\to\rb$ is a morphism between
two $C^*$-algebras then $\varphi\mapsto\cp(\varphi(H))$ is an
observable affiliated to $\rb$ denoted $\cp(H)$. So
$\cp(\varphi(H))=\varphi(\cp(H))$. If $\ra$ and $\rb$ are realized
on Hilbert spaces $\ch_a,\ch_b$, then any self-adjoint operator $H$
on $\ch_a$ affiliated to $\ra$ defines an observable affiliated to
$\ra$, but the observable $\cp(H)$ is not necessarily associated to
a self-adjoint operator on $\ch_b$ because the natural operator
associated to it could be non-densely defined (in our context, it
often has domain equal to $\{0\}$). The spectrum and essential
spectrum of an observable are defined in an obvious way \cite{ABG}. 

Now clearly, if $H$ is an observable affiliated to $\re(X)$ then
$\vkappa.H$ defined by $\varphi(\vkappa.H)=\vkappa.\varphi(H)$ is an
observable affiliated to $\re_\vkappa(X)$. This is the
\emph{localization of $H$ at} $\vkappa$ and we have
\begin{equation}\label{eq:esss}
\spec_\ess(H)=\overline{\ccup}_{\vkappa\in\delta(X)}\spec(\vkappa.H).
\end{equation}

We shall not give in this paper affiliation criteria specific to the
algebra $\re(X)$ but the results of Section \ref{s:groups} and the
examples form \cite{GI} should convince the reader that the class of
operators affiliated to $\re(X)$ is very large. On the other hand,
if $H$ is a positive self-adjoint operator such that
$\rme^{-H}\in\re(X)$ then $H$ is affiliated to $\re(X)$. Or this is
condition is certainly satisfied by the Laplace operator associated
to a large class of Riemannian manifolds due to known estimates on
the heat kernel of the manifold. We thank Thierry Coulhon for an
e-mail exchange on this question.  

In connection with Proposition \ref{pr:evk} we mention that in
Section \ref{s:ide} we consider a second class of ideals
$\rg_\vkappa(X)$ in $\re(X)$ which are similar to the
$\re_{(\vkappa)}(X)$. More precisely, let $\rg_\vkappa(X)$ be
defined as the right hand side of \eqref{eq:ek} for any
$\vkappa\in\delta(X)$. Then $\rg_\vkappa(X)$ is an ideal of $\re(X)$
and $\re_{(\vkappa)}(X)\subset\rg_\vkappa(X)$ where equality holds
if $X$ is a space of class A but the inclusion is strict in general.
We say that \emph{$\rg_\vkappa$ is the ghost envelope of
  $\re_{(\vkappa)}$}. Thus \emph{for each ultrafilter
  $\vkappa\in\delta(X)$ we may have two distinct contributions to
  the essential spectrum of $H$ associated to $\vkappa$: first the
  spectrum of the localization $\vkappa.H=H/\re_{(\vkappa)}$ at
  $\vkappa$ and second the spectrum of $H/\rg_{\vkappa}$, which is a
  subset of the first one.}

In particular, besides the smallest ideal $\rk(X)$ of $\re(X)$ there
is a second ``small'' ideal which appears quite naturally in the
theory. This is the \emph{ghost ideal} defined by
\begin{equation}\label{eq:phantom}
  \rg(X)=
  \{T\in\re(X) \mid \lim_{x\to\infty}\|\ind_{B_x(r)}T\|=0 \text{ for
    all } r>0 \}. 
\end{equation}
The operators $T\in\rg(X)$ vanish everywhere at infinity in the
configuration space $X$ but could be not compact. The role of the
Property A is to ensure that $\rg(X)=\rk(X)$. For discrete metric
spaces of bounded geometry, this phenomenon is studied in detail by
Chen and Wang, see \cite{CW1,CW2,Wa} and references
therein. Proposition \ref{pr:discrete} shows, among other things,
that our definition of the ghost ideal in the discrete case
coincides with theirs.

Observe that in general, if $H$ is an observable affiliated to
$\re(X)$ then the \emph{ghost spectrum of $H$}, i.e.\ the spectrum
of the quotient observable $H/\rg(X)$, is strictly included in the
essential spectrum of $H$.

\section{The elliptic $C^*$-algebra} \label{s:ell}

In this section $X=(X,d,\mu)$ is a metric space $(X,d)$ equipped
with a measure $\mu$ and such that:

\begin{itemize}
\item $(X,d)$ is a locally compact not compact metric space and each
  closed ball is a compact set,
\item $\mu$ is a Radon measure on $X$ with support equal to $X$ and 
$\sup_x\mu(B_x(r))=V(r)<\infty\;\forall r>0$.
\end{itemize}

If $k$ is a controlled kernel let $d(k)$ be the least number $r$
such that $d(x,y)>r \Rightarrow k(x,y)=0$. Recall that
\begin{equation}\label{eq:frk}
\ccfr(X^2)=
\{ k: X^2\to\C \mid k \text{ is a bounded uniformly
  continuous  controlled kernel} \}.
\end{equation} 
If $k\in\ccfr(X^2)$ then $Op(k)$ is the  operator on $L^2(X)$ given
by $(Op(k)f)(x)=\int_X k(x,y) f(y) \d x$. From
\begin{equation}\label{eq:schur}
\|Op(k)\|^2\leq \sup_x\int |k(x,y)| \d y \cdot
\sup_y\int |k(x,y)| \d x,
\end{equation}
which is the Schur estimate, we get
\begin{equation}\label{eq:shur}
\|Op(k)\| \leq V(d(k))\sup|k|.
\end{equation} 
If $k,l\in\ccfr(X^2)$ then we denote $k^*(x,y)=\bar{k}(y,x)$ and
$(k\star l)(x,y)=\int k(x,z)l(z,y) \d z$. Clearly $Op(k)^*=Op(k^*)$
and $Op(k) Op(l)= Op(k\star l)$. The following simple fact is
useful.

\begin{lemma}\label{lm:kl}
  If $k,l\in\ccfr(X^2)$ then $k\star l\in\ccfr(X^2)$, we have
  $d(k\star l)\leq d(k)+d(l)$, and
\[
\sup|k\star l|\leq \sup|k| \cdot \sup|l| 
\cdot \min\{V(d(k)), V(d(l))\}.
\]
\end{lemma}
\proof
If we set $s=d(k)$ and $t=d(l)$ then clearly
\[
|(k\star l)(x,y)| \leq \sup|k| \cdot \sup|l|\cdot 
\mu \left( B_x(s)\cap B_y(t) \right)
\]
which gives both estimates from the statement of the lemma. 
To prove the uniform continuity we use
\begin{align*}
|(k\star l)(x,y) - (k\star l)(x',y)|  & \leq
\sup_z |k(x,z) - k(x',z)| \int |l(z,y)| \d z  \\
& \leq 
\sup_z |k(x,z) - k(x',z)| \cdot \sup|l| \cdot V(t)
\end{align*} 
and a similar inequality for $|(k\star l)(x,y) - (k\star l)(x,y')|$.
\qed 

Thus $\ccfr(X^2)$, when equipped with the usual linear structure and
the operations $k^*$ and $k\star l$, becomes a $*$-algebra and
$k\mapsto Op(k)$ is a morphism into $\rb(X)$ hence its range is a
$*$-subalgebra of $\rb(X)$. Hence the elliptic algebra $\re(X)$
defined in \eqref{eq:ell} is a $C^*$-algebra of operators on
$L^2(X)$.

The uniform continuity assumption involved in the definition
\eqref{eq:frk} of $\ccfr(X)$ hence in that of $\re(X)$ is important
because thanks to it we have $\re(X)=\cc(X)\rtimes_\rmr X$ if $X$ is
a unimodular locally compact group, cf.\ Sections \ref{s:groups} and
\ref{s:psd}. Here $\cc(X)$ is the $C^*$-algebra of right uniformly
continuous functions on $X$ on which $X$ acts by left translations
and $\rtimes_\rmr$ denotes the reduced crossed product. In
particular, the equality $\cc(X)\rtimes_\rmr X = \re(X)$ gives a
description of the crossed product independent of the group
structure of $X$.

We say that $T\in\rb(X)$ is a \emph{controlled operator} if there is
$r>0$ such that if $F,G$ are closed subsets of $X$ with $d(F,G)>r$
then $\ind_F T \ind_G=0$; let $d(T)$ be the smallest $r$ for which
this holds (see \cite{R}; this class of operators has also been
considered in \cite{D} and in \cite{GG}).  Observe that the $Op(k)$
with $k\in\ccfr(X^2)$ are controlled operators but if $X$ is not
discrete then there are many others and most of them do not belong
to $\re(X)$.  The norm closure of the set of controlled operators
will be discussed in Section \ref{s:psd}.

Since the kernel of $\varphi(Q) Op(k)$ is $\varphi(x)k(x,y)$ and that
of $ Op(k)\varphi(Q)$ is $k(x,y)\varphi(y)$, we clearly have
\[
\cc(X)\re(X)=\re(X)\cc(X)=\re(X).
\]
This defines a $\cc(X)$-bimodule structure on $\re(X)$. We note
that, as a consequence of the Cohen-Hewitt theorem,
\emph{if $\ca$ is a $C^*$-subalgebra of $\cc(X)$ then the set
$\ca\ce(X)$ consisting of products $AT$ of elements $A\in\ca$ and
$T\in\re(X)$ is equal to the closed linear subspace of $\re(X)$
generated by these products.}

\begin{proposition}\label{pr:kx}
We have $\rk(X)=\Co(X)\re(X)=\re(X)\Co(X)\subset\re(X)$. 
\end{proposition}
\proof If $\varphi\in\Cc$ and $k\in\ccfr$ then the operator
$\varphi Op(k)$ has kernel $\varphi(x)k(x,y)$ which is a continuous
function with compact support on $X^2$, hence $\varphi Op(k)$ is a
Hilbert-Schmidt operator. Thus we have $\Co(X)\re(X)\subset\rk(X)$
and by taking adjoints we also get
$\re(X)\Co(X)\subset\rk(X)$. Conversely, an operator with kernel in
$\Cc(X^2)$ clearly belongs to $\Cc(X)\re(X)$ for example.  \qed

$\re(X)$ is a non-degenerate $\Co(X)$-bimodule and there is a
natural topology associated to such a structure, we call it the
local topology on $\re(X)$. Its utility will be clear from Section
\ref{s:groups}.

\begin{definition}\label{df:local}
  The \emph{local topology} on $\re(X)$ is the topology associated
  to the family of seminorms
  $\|T\|_\theta=\|T\theta(Q)\|+\|\theta(Q)T\|$ with
  $\theta\in\Co(X)$.
\end{definition}

This is the analog of the topology of local uniform convergence on
$\cc(X)$.  Obviously one may replace the $\theta$ with
$\ind_\Lambda$ where $\Lambda$ runs over the set of compact subsets
of $X$.  If $T\in\re(X)$ and $\{T_\alpha\}$ is a net of operators in
$\re(X)$ we write $T_\alpha\to T$ or $\lim_\alpha T_\alpha=T$
\emph{locally} if the convergence takes place in the local topology.
Since $X$ is $\sigma$-compact there is $\theta\in\Co(X)$ with
$\theta(x)> 0$ for all $x\in X$ and then $\|\cdot\|_\theta$ is a
norm on $\re(X)$ which induces on bounded subsets of $\re(X)$ the
local topology.

The local topology is finer than the $*$-strong operator topology
inherited from the embedding $\re(X) \subset \rb(X)$.  We may also
consider on $\re(X)$ the (intrinsically defined) strict topology
associated to the smallest essential ideal $\rk(X)$; this is weaker
than the local topology and finer than the $*$-strong operator
topology, but coincides with the last one on bounded sets.

\begin{lemma}\label{lm:loctop}
The involution $T\mapsto T^*$ is locally continuous on $\re(X)$. The
multiplication is locally continuous on bounded sets.
\end{lemma}
\proof Since $\|T^*\|_\theta=\|T\|_{\bar{\theta}}$ the first
assertion is clear. Now assume $S_\alpha\to S$ locally and
$\|S_\alpha\|\leq C$ and $T_\alpha\to T$ locally. If $\theta\in\Co$
then $T\theta$ is a compact operator so there is $\theta'\in\Co$
such that $T\theta=\theta'K$ for some compact operator $K$. Then we
write $ (S_\alpha T_\alpha - S T)\theta = S_\alpha (T_\alpha -
T)\theta + (S_\alpha - S )\theta' K $. \qed

The \emph{ghost ideal} is defined as follows:
\begin{equation}\label{eq:ghost}
\rg(X):=
\{T\in\re(X) \mid \lim_{x\to\infty}\|\ind_{B_x(r)}T\|=0 \;\forall r \}
=\{T\in\re(X) \mid \lim_{x\to\infty}\|T\ind_{B_x(r)}\|=0 \;\forall r \}. 
\end{equation}
The fact that $\rg$ is an ideal of $\re$ follows from the equality
stated above which in turn is proved as follows: for each
$\varepsilon>0$ there is a controlled kernel $k$ such that
$\|T-Op(k)\|<\varepsilon$ hence if $R=r+d(k)$ we have
$$
\|T\ind_{B_x(r)}\| <\varepsilon + \|Op(k) \ind_{B_x(r)}\|=
\varepsilon + \|\ind_{B_x(R)}Op(k) \ind_{B_x(r)}\|
< 2\varepsilon + \|\ind_{B_x(R)}T\|
$$
which is less than $3\varepsilon $ for large $x$.

We have $\rk(X)\subset\rg(X)$ because
$\lim_{x\to\infty}\ind_{B_x(r)}=0$ strongly on $L^2$.  It is known
that \emph{the inclusion is strictly in general}
\cite[p.\,349]{HLS}. In the rest of this section we prove that
equality holds if $X$ is of class A. We begin with some general
useful remarks.

\begin{lemma}\label{lm:capacity}
  If \eqref{eq:12} holds then there a subset $Z\subset X$   with 
  $X=\cup_{z\in Z}B_z$ and a function $N:\R\to\mbn$ such that:  
  for any $x\in X$ and  $r\geq 1$     the number of $z\in Z$ such that
  $B_z(r)\cap B_x(r)\neq\emptyset$ is at most $N(r)$.
\end{lemma}
\proof Let $Z$ be a maximal subset of $X$ such that $d(a,b)>1$ if
$a,b$ are distinct points in $Z$. Then we have $X=\cup_{z\in Z}B_z$
(the contrary would contradict the maximality of $Z$).  Now fix
$r\geq 1$, let $x\in X$, denote $Z_x$ the set of $z\in Z$ such that
$B_z(r)\cap B_x(r)\neq\emptyset$, and let $N_x$ be the number of
elements of $Z_x$.  Choose $a\in Z$ such that $x\in B_a$. Then
$B_x(r)\subset B_a(r+1)$ hence if $z\in Z_x$ then $B_z(r)\cap
B_a(r+1)\neq\emptyset$ so $d(z,a)\leq 2r+1$.  Since the balls
$B_z(1/2)$ corresponding to these $z$ are pairwise disjoint and
included in $B_a(2r+2)$, the volume of their union is larger than
$\nu N_x$, where $\nu=\inf_{y\in X}\mu(B_y(1/2))$, and smaller than
$V(2r+2)$, hence $N_x\leq V(2r+2)/\nu$. Thus we may take
$N(r)=V(2r+2)/\nu$.  \qed

From now on, if \eqref{eq:12} is satisfied, the set $Z$ and the
function $N$ will be as in Lemma \ref{lm:capacity}.

\begin{lemma}\label{lm:estime}
If \eqref{eq:12} is satisfied and $T$ is a controlled operator, then
\begin{equation}\label{eq:estime}
\|T\|\leq N(d(T)+1)^{1/2}\sup_{x\in X} \|\ind_{B_x} T\|. 
\end{equation}
\end{lemma}
\proof
Set $R=d(T)+1$. Then for any $f\in L^2$ we have
$$
\|Tf\|^2\leq \sum_{z\in Z}\|\ind_{B_z}Tf\|^2 = 
\sum_{z\in Z}\|\ind_{B_z}T \ind_{B_z(R)}f\|^2  \leq
\sup_{z\in Z}\|\ind_{B_z}T\|^2 \sum_{z\in Z}\|\ind_{B_z(R)}f\|^2
$$
and from Lemma \ref{lm:capacity} we get 
$\sum_{z\in Z}\ind_{B_z(R)}\leq N(R)$.
\qed

\begin{lemma}\label{lm:capc}
  Assume that \eqref{eq:12} is satisfied and let $T\in\rb(X)$. If
  $\lim_{x\to\infty}\|\ind_{B_x(r)}T\|=0$ holds for $r=1$ then it
  holds for all $r>0$. In particular, we have
\begin{equation}\label{eq:ghosts}
\rg(X)=
\{T\in\re(X) \mid \lim_{x\to\infty}\|\ind_{B_x}T\|=0\}
=\{T\in\re(X) \mid \lim_{x\to\infty}\|T\ind_{B_x}\|=0\}. 
\end{equation}
\end{lemma}
\proof Let $r>1$, $\varepsilon>0$ and let $F$ be a finite subset of
$Z$ such that $\|\ind_{B_z}T\|<\varepsilon/N(r)$ if $z\in Z
\setminus F$. We consider points $x$ such that $d(x,F)>r+1$ and
denote $Z(x,r)$ the set of $z\in Z$ such that $B_z\cap
B_x(r)\neq\emptyset$. Then $Z(x,r)$ has at most $N(r)$ elements and
$B_x(r)\subset\cup_{z\in Z(x,r)}B_z$ hence $\|\ind_{B_x(r)}T\|\leq
N(r) \max_{z\in Z(x,r)}\|\ind_{B_z}T\|<\varepsilon$ because $F\cap
Z(x,r)=\emptyset$.  \qed

An operator $T\in\rb(X)$ is called \emph{locally compact} if for any
compact set $K$ the operators $\ind_KT$ and $T\ind_K$ are
compact. Clearly \emph{any operator in $\re(X)$ is locally compact}.

\begin{lemma}\label{lm:clcg}
  Assume that \eqref{eq:12} is satisfied. If $T\in\rb(X)$ is a
  controlled locally compact operator such that
  $\|\ind_{B_x}T\|\to0$ as $x\to\infty$ then $T$ is compact.
\end{lemma}
\proof 
Choose $o\in X$ and let $\ind_R$ be the characteristic function of
the ball $B_o(R)$. Then $\ind_R T$ is compact so it suffices to show
that $\ind_R T$ converges in norm to $T$ as $R\to\infty$. Clearly
$T-\ind_R T$ is controlled with $d(T-\ind_R T)\leq d(T)$ hence from
Lemma \ref{lm:estime} we get 
$$
\|T-\ind_R T\|\leq C \sup_{x\in X}\|\ind_{B_x}(\ind-\ind_R)T\|
\leq C \sup_{d(x,o)>R-1}\|\ind_{B_x}T\|
$$ 
which proves the lemma. \qed

%Assume that $T\in\rb(X)$ is locally compact and has the
%property $\ind_F T \ind_G=0$ if $F,G\subset X$ satisfy $d(F,G)>r$
%for some fixed $r$.  To prove the compactness of $T$ it suffices to
%show that $\|\ind_R T\|\to0$ as $R\to\infty$, where $\ind_R$ is the
%characteristic function of the set of points $x$ such that
%$d(x,o)>R$ for some fixed $o\in X$. We set $|x|=d(x,o)$ and below
%denote by $z$ points in $Z$. Then
%$$
%\|\ind_{R+1}Tf\|^2\leq \sum_{|z|>R}\|\ind_{B_z}Tf\|^2 = 
%\sum_{|z|>R}\|\ind_{B_z}T \ind_{B_z(r+1)}f\|^2  \leq
%\sup_{|z|>R}\|\ind_{B_z}T\|^2 \sum_{z}\|\ind_{B_z(r+1)}f\|^2
%$$
%and the last sum is $\leq C(r)^2\|f\|^2$ by Lemma
%\ref{lm:capacity}. Thus $\|\ind_{R+1}T\|\leq
%C(r)\sup_{|z|>R}\|\ind_{B_z}T\|$.  

Now we use an idea from \cite{CW1} (truncation of kernels with the
help of functions of positive type) and the technique of the proof
of Theorem 5.1 from \cite{Pi}.

Let $\ch$ be an arbitrary separable Hilbert space (in Definition
\ref{df:yu} we took $\ch=L^2(X)$) and let $\phi:X\to\ch$ be a Borel
function such that $\|\phi(x)\|=1$ for all $x$. Define
$M_\phi:L^2(X)\to L^2(X;\ch)=L^2(X)\otimes \ch$ by $(M_\phi
f)(x)=f(x)\phi(x)$. Then $M_\phi$ is a linear operator with
$\|M_\phi\|=1$ and its adjoint $M_\phi^*:L^2(X;\ch)\to L^2(X)$ acts
as follows: $(M_\phi^*F)(x)=\braket{\phi(x)}{F(x)}$.  Let $T\mapsto
T_\phi$ be the linear continuous map on $\rb(X)$ given by
$T_\phi=M_\phi^*(T\otimes1)M_\phi$. Clearly $\|T_\phi\|\leq\|T\|$.

Let $k:X^2\to\C$ be a locally integrable function. We say that an
operator $T\in\rb(X)$ has integral kernel $k$ if
$\braket{f}{Tg}=\int_{X^2} k(x,y) \bar{f}(x)g(y) \d x \d y$ for all
$f,g\in\Cc(X)$. If $k$ is a Schur kernel, i.e.\ $\sup_x\int_X
(|k(x,y)|+|k(y,x)|)\d y<\infty $, then we say that $T$ is a Schur
operator and we have the estimate \eqref{eq:schur} for its norm. And
$T$ is a Hilbert-Schmidt operator if and only if $k\in
L^2(X^2)$. From the relation $\braket{f}{T_\phi
  g}=\braket{f\phi}{T\otimes1 g\phi}$ valid for $f,g\in\Cc(X)$ we
easily get:

\begin{lemma}\label{lm:fkern}
  If $T$ has kernel $k$ then $T_\phi$ has kernel
  $k_\phi(x,y)=\braket{\phi(x)}{\phi(y)}k(x,y)$. In particular, if
  $T$ is a Schur, Hilbert-Schmidt, or compact operator, then
  $T_\phi$ has the same property.
\end{lemma}

\begin{lemma}\label{lm:cfkern}
  Assume that $\braket{\phi(x)}{\phi(y)}=0$ if $d(x,y)>r$. Then for
  each $T\in\rb(X)$ the operator $T_\phi$ is controlled, more
  precisely: if $F,G$ are closed subsets of $X$ with $d(F,G)>r$ then
  $\ind_F T_\phi \ind_G=0$.
\end{lemma}
\proof We have to prove that $\braket{\ind_F f}{T_\phi \ind_G g}=0$
for all $f,g\in L^2(X)$ and $T\in\rb(X)$. The map $T\mapsto T_\phi$
is continuous for the weak operator topology and the set of finite
range operators is dense in $\rb(X)$ for this topology. Thus it
suffices to assume that $T$ is Hilbert-Schmidt (or even of rank one)
and then the assertion is clear by Lemma \ref{lm:fkern}.  \qed

Observe that if $\theta:X\to\C$ is a bounded Borel function then
$M_\phi\theta(Q)= (\theta(Q)\otimes1)M_\phi$ hence $\theta
T_\phi=(\theta T)_\phi$ and $T_\phi\theta =(T\theta)_\phi$ with the
usual abbreviation $\theta=\theta(Q)$. In particular, Lemma
\ref{lm:fkern} implies:

\begin{lemma}\label{lm:lc}
  Let $T\in\rb(X)$. If $T$ is locally compact then $T_\phi$ is
  locally compact. If $\|\ind_{B_x(r)}T\|\to0$ as $x\to\infty$, then
  $\|\ind_{B_x(r)}T_\phi\|\to0$ as $x\to\infty$.
\end{lemma}

\begin{theorem}\label{th:gc}
  If $X$ is a class A space then $\rk(X)=\rg(X)$.
\end{theorem}
\proof Let $T\in\rg(X)$ and $\phi$ as above. Then $T$ is locally
compact hence $T_\phi$ is locally compact, and we have
$\|\ind_{B_x}T_\phi\|\to0$ as $x\to\infty$ by Lemma
\ref{lm:lc}. Moreover, if $\phi$ is as in Lemma \ref{lm:cfkern} then
$T_\phi$ is controlled so, by Lemma \ref{lm:clcg}, $T_\phi$ is
compact. Thus it suffices to show that any $T\in\re(X)$ is a norm
limit of operators $T_\phi$ with $\phi$ of the preceding form. Since
$T\mapsto T_\phi$ is a linear contraction, it suffices to show this
for operators of the form $T=Op(k)$ with $k\in\ccfr(X^2)$. But then
$T-T_\phi$ is an operator with kernel
$k(x,y)(1-\braket{\phi(x)}{\phi(y)})$ hence, if we denote
$M=\sup|k|, d=d(k)$, from \eqref{eq:schur} we get
$$
\|T-T_\phi\|\leq M \sup_x \int_{B_x(d)}
|1-\braket{\phi(x)}{\phi(y)}| \d y.
$$
Until now we did not use the fact that $\ch=L^2(X)$ in Definition
\ref{df:yu}. If we are in this situation note that we may replace
$\phi(x)$ by $|\phi(x)|$ and then $\braket{\phi(x)}{\phi(y)}$ is
real. More generally, assume that the $\phi(x)$ belong to a real
subspace of the (abstract) Hilbert space $\ch$ so that
$\braket{\phi(x)}{\phi(y)}$ is real for all $x,y$. Then
$1-\braket{\phi(x)}{\phi(y)}=\|\phi(x)-\phi(y)\|^2/2$ so we have
$$
\|T-T_\phi\|\leq (M/2)\sup_x \int_{B_x(d)} \|\phi(x)-\phi(y)\|^2 \d y.
$$
Since $X$ has Property A, one may choose $\phi$ such that this be
smaller than any given number.  \qed

\section{Coarse filters on $X$ and ideals of $\cc(X)$} \label{s:FF}

\subsection{Filters}\label{s:filt}
We recall some elementary facts; for the moment $X$ is an arbitrary
set. A \emph{filter} on $X$ is a nonempty set $\xi$ of subsets
of $X$ which is stable under finite intersections, does not contain
the empty set, and has the property: $G\supset F\in\xi
\Rightarrow G\in\xi$. If $Y$ is a topological space and
$\phi:X\to Y$ then $\lim_\xi\phi=y$ or
$\lim_{x\to\xi}\phi(x)=y$ means that $y\in Y$ and if $V$ is a
neighborhood of $y$ then $\phi^{-1}(V)\in\xi$.

The set of filters on $X$ is equipped with the order relation given
by inclusion. Then the trivial filter $\{X\}$ is the smallest filter
and the lower bound of any nonempty set $\cf$ of filters exists:
$\inf\cf=\cap_{\xi\in\cf}\xi$. A set $\cf$ of filters is
called \emph{admissible} if
$\cap_{\xi\in\cf}F_\xi\neq\emptyset$ if
$F_\xi\in\xi$ for all $\xi$ and $F_\xi=X$ but for a
finite number of indices $\xi$. If $\cf$ is admissible then the
upper bound $\sup\cf$ exists: this is the set of sets of the form
$\cap_{\xi\in\cf}F_\xi$ where $F_\xi\in\xi$ for all
$\xi$ and $F_\xi=X$ but for a finite number of indices
$\xi$.

Let $\beta(X)$ be the set of ultrafilters on $X$. If $\xi$ is a
filter let $\xi^\dagger$ be the set of ultrafilters finer than
it. Then $\xi=\inf\xi^\dagger$. We equip $\beta(X)$ with the
topology defined by the condition: a nonempty subset of $\beta(X)$
is closed if and only if it is of the form $\xi^\dagger$ for some
filter $\xi$. Note that for the trivial filter consisting of only
one set we have $\{X\}^\dagger=\beta(X)$.  Then $\beta(X)$ becomes a
compact topological space, this is the Stone-\v{C}ech
compactification of the \emph{discrete} space $X$, and is naturally
identified with the spectrum of the $C^*$-algebra of all bounded
complex functions on $X$. There is an obvious dense embedding
$X\subset\beta(X)$, any bounded function $\varphi:X\to\C$ has a
unique continuous extension $\beta(\varphi)$ to $\beta(X)$, and any
map $\phi:X\to X$ has a unique extension to a continuous map
$\beta(\phi):\beta(X)\to\beta(X)$. 

More generally, if $Y$ is a compact topological space, each map
$\phi:X\to Y$ has a unique extension to a continuous map
$\beta(\phi):\beta(X)\to Y$.  The following simple fact should be
noticed: if $\xi$ is a filter and $o$ is a point in $Y$ then
$\lim_\xi\phi=o$ is equivalent to
$\beta(\phi)|\xi^\dagger=o$.\label{p:fil} Indeed, $\lim_\xi\phi=o$
is equivalent to $\lim_\vkappa\phi=o$ for any
$\vkappa\in\xi^\dagger$ (for the proof, observe that if this last
relation holds then for each neighborhood $V$ of $o$ the set
$\phi^{-1}(V)$ belongs to $\vkappa$ for all $\vkappa\in\xi^\dagger$,
hence $\phi^{-1}(V)\subset\cap_{\vkappa\in\xi^\dagger}\vkappa=\xi$).

Now assume that $X$ is a locally compact non-compact topological
space.  Then the \emph{Fr\'echet filter} is the set of complements
of relatively compact sets; we denote it $\infty$, so that
$\lim_{x\to\infty}\phi(x)=y$ has the standard meaning. Let
$\delta(X)=\infty^\dagger$ be the set of ultrafilters finer than it.
Thus $\delta(X)$ is a compact subset of $\beta(X)$ and we have
$\delta(X)\subset \beta(X)\setminus X$ (strictly in general):
\[
\delta(X)=\{ \vkappa\in\beta(X) \mid 
\text{ if } K \subset X\text{ is relatively compact then }
K\nin\vkappa \}. 
\]
Indeed, if $\vkappa$ is an ultrafilter then for any set $K$ either
$K\in\vkappa$ or $K^\c\in\vkappa$. If we interpret $\vkappa$ as a
character of $\ell_\infty(X)$ then $\vkappa\in\delta(X)$ means
$\vkappa(\varphi)=0$ for all $\varphi\in\Co(X)$.

\subsection{Coarse filters}\label{s:cfilt}
Now assume that $X$ is a metric space. If $F\subset X$ then
$\bar{F}$ is its closure and $F^\c =X\setminus F$ its complement.
We set $d_F(x):=\inf_{y\in F} d(x,y)$.  Note that $d_F=d_{\bar{F}}$
and $|d_F(x)-d_F(y)|\leq d(x,y)$.  If $r>0$ let $F_{(r)}:=\{x\mid
d(x,F)\leq r\}=\cup_{x\in F} B_x(r)$ be the neighborhood ``of order
$r$'' of $F$.

If $r>0$ we denote $F^{(r)}$ the set of points $x$ such that
$d(x,F^\c)>r$. This is an open subset of $X$ included in $F$ and at
distance $r$ from the boundary of $F$ (so if $F$ is too thin,
$F^{(r)}$ is empty). In other terms, $x\in F^{(r)}$ means that there
is $r'>r$ such that $B_x(r')\subset F$. In particular,
$(F^{(r)})_{(r)}\subset F$ and for an arbitrary pair of sets $F,G$
we have $(F\cap G)^{(r)}= F^{(r)}\cap G^{(r)}$ and $F\subset G
\Rightarrow F^{(r)}\subset G^{(r)}$. \label{p:fg}

We say that a filter $\xi$ is \emph{coarse} if for any $F\in\xi$ and
$r>0$ we have $F^{(r)}\in\xi$.  We emphasize that this should hold
for \emph{all} $r>0$. If for each $F\in\xi$ there is $r>0$ such that
$F^{(r)}\in\xi$ then the filter is called \emph{round}.
Equivalently, $\xi$ is coarse if for each $F\in\xi$ and $r>0$ there
is $G\in\xi$ such that $G_{(r)}\subset F$ and $\xi$ is round if for
each $F\in\xi$ there are $G\in\xi$ and $r>0$ such that
$G_{(r)}\subset F$.

Our terminology is related to the notion of coarse ideal introduced
in \cite{HPR} (our space $X$ being equipped with the bounded metric
coarse structure). More precisely, a \emph{coarse ideal} is a set
$\ci$ of subsets of $X$ such that $B\subset A\in\ci\Rightarrow
B\in\ci$ and $A\in\ci\Rightarrow A_{(r)}\in\ci$ for all $r>0$.
Clearly $\ci\mapsto\ci^\c:=\{A^\c \mid A\in\ci\}$ is a one-one
correspondence between coarse ideals and filters.

Coarse filters on groups are very natural objects: \emph{if $X$ is a
  group, then a round filter is coarse if and only if it is
  translation invariant} (Proposition \ref{pr:gcoarse}).

The Fr\'echet filter is coarse because if $K$ is relatively compact
then $K_{(r)}$ is compact for any $r$ (the function $d_K$ is proper
under our assumptions on $X$). The trivial filter $\{X\}$ is coarse.

More general examples of coarse filters are constructed as follows
\cite{D,GI0}.  Let $L\subset X$ be a set such that $L_{(r)}\neq X$
for all $r>0$.  Then the filter generated by the sets
$L_{(r)}^c=\{x\mid d(x,L)>r\}$ when $r$ runs over the set of
positive real numbers is coarse (indeed, it is clear that the
$L_{(r)}$ generate a coarse ideal). If $L$ is compact the associated
filter is $\infty$. If $X=\R$ and $L=]-\infty,0]$ then the
corresponding filter consists of neighborhoods of $+\infty$ and this
example has obvious $n$-dimensional versions. If $L$ is a sparse set
(i.e. the distance between $a\in L$ and $L\setminus\{a\}$ tends to
infinity as $a\to\infty$) then the ideal in $\cc(X)$ associated to
it (cf. below) and its crossed product by the action of $X$ (if $X$
is a group) are quite remarkable objects, cf. \cite{GI0}.  It should
be clear however that most coarse filters are not associated to any
set $L$.

Let $X$ be an Euclidean space and let $G(X)$ be the set of finite
unions of strict vector subspaces of $X$. The sets $L_{(r)}^\c$ when
$L$ runs over $G(X)$ and $r$ over $\R_+$ form a filter basis and the
filter generated by it is the \emph{Grassmann filter} $\gamma$ of
$X$.  This is a translation invariant hence coarse filter which
plays a role in a general version of the $N$-body problem, see
\cite[Section 6.5]{GI}. The relation $\lim_\gamma\varphi=0$ means
that the function $\varphi$ vanishes when we are far from any strict
affine subspace.

\begin{lemma}\label{lm:st}
  If $\cf$ is a nonempty set of coarse filters then $\inf\cf$ is a
  coarse filter. If $\cf$ is admissible then $\sup\cf$ is a coarse
  filter.
\end{lemma}
\proof If $F\in\inf\cf=\cap_{\xi\in\cf}\xi$ then for any $r>0$ and
$\xi$ we have $F^{(r)}\in\xi$ and so
$F^{(r)}\in\cap_{\xi\in\cf}\xi$. Now assume for example that
$F\in\xi$ and $G\in\eta$ with $\xi,\eta\in\cf$ and let $r>0$. Then
there are $F'\in\xi$ and $G'\in\eta$ such that $F'_{(r)}\subset F$
and $G'_{(r)}\subset G$ hence $(F'\cap G')_{(r)}\subset F'_{(r)}
\cap G'_{(r)}\subset F\cap G$.  The argument for sets of the form
$\cap_\xi F_\xi$ with $F_\xi=X$ but for a finite number of indices
$\xi$ is similar.  \qed

\begin{lemma}\label{lm:coarsefr}
  A coarse filter is either trivial, and then
  $\xi^\dagger=\beta(X)$, or finer than the Fr\'echet filter, and
  then $\xi^\dagger\subset\delta(X)$.
\end{lemma}
\proof Assume that $\xi$ is not finer than the Fr\'echet
filter. Then there is a compact set $K$ such that
$K^\c\notin\xi$. Hence for any $F\in\xi$ we have $F\not\subset K^\c$
so $F\cap K\neq\emptyset$. Note that the closed sets in $\xi$ form a
basis of $\xi$ (if $F\in\xi$ then the closure of $F^{(2)}$ belongs
to $\xi$ and is included in $F^{(1)}$ hence in $F$). The set
$\{F\cap K \mid F\in\xi \text{ and is closed}\}$ is a filter basis
consisting of closed sets in the compact set $K$ hence there is
$a\in K$ such that $a\in F$ for all $F\in\xi$. Then if $F\in\xi$ and
$r>0$ there is $G\in\xi$ such that $G_{(r)}\subset F$ and since
$a\in G$ we have $B_a(r)\subset G_{(r)}\subset F$. But $X=\cup_r
B_a(r)$ so $X\subset F$.  \qed

\subsection{Coarse ideals of $\cc(X)$}\label{s:ideals}

We now recall some facts concerning the relation between filters on
$X$ and ideals of $\cc(X)$. To each filter $\xi$ on $X$ we
associate an ideal $\ci_\xi(X)$ of $\cc(X)$:
\begin{equation}\label{eq:idfil}
  \ci_\xi(X) := \{ \varphi\in\cc(X) \mid \lim_\xi
  \varphi=0\} 
\end{equation}
If $\xi$ is the Fr\'echet filter then $\lim_\xi\varphi=0$
means $\lim_{x\to\infty}\varphi(x)=0$ in the usual sense and so the
corresponding ideal is $\Co(X)$. The ideal associated to the trivial
filter clearly is $\{0\}$. We also have:
\begin{align}
& \xi\subset\eta\Rightarrow 
\ci_{\xi}(X)\subset\ci_{\eta}(X) \label{eq:sub}\\
& \ci_{\xi\cap\eta}(X) \label{eq:int}=
\ci_{\xi}(X)\cap\ci_{\eta}(X)=
\ci_{\xi}(X)\ci_{\eta}(X)
\end{align}
The \emph{round envelope} $\xi^\circ$ of $\xi$ is the finer
round filter included in $\xi$. Clearly this is the filter
generated by the sets $F_{(r)}$ when $F$ runs over $\xi$ and $r$
over $\R_+$. Note that
$\ci_\xi(X)=\ci_{\xi^\circ}(X)$, i.e.  for
$\varphi\in\cc(X)$ we have $\lim_\xi\varphi=0$ if and only if
$\lim_{\xi^\circ}\varphi=0$. Indeed, if $\varepsilon>0$ let $F$
be the set of points were $|\varphi(x)|<\varepsilon/2$ and let $r>0$
be such that $|\varphi(x)-\varphi(y)|<\varepsilon/2$ if $d(x,y)\leq
r$.  Then $|\varphi(x)|<\varepsilon$ if $x\in F_{(r)}$.

We recall a well-known description of the spectrum of the algebra
$\cc(X)$ in terms of round filters.

\begin{proposition}\label{pr:sam}
  The map $\xi\mapsto\ci_\xi(X)$ is a bijection between
  the set of all round filters on $X$ and the set of all ideals of
  $\cc(X)$.
\end{proposition}

An ideal $\ci$ of $\cc(X)$ will be called \emph{coarse} if for each
positive $\varphi\in\ci$ and $r>0$ there is a positive $\psi\in\ci$
such that
\begin{equation}\label{eq:fid}
d(x,y)\leq r \text{ and } \psi(y)<1 \Rightarrow \varphi(x)<1.
\end{equation}

\begin{lemma}\label{lm:dfg} 
Let $F,G$ be subsets of $X$ such that $G_{(r)}\subset F$. Then the
function $\theta= d_{F^\c} \left( d_{F^\c} +d_G \right)^{-1}$
belongs to $\cc(X)$  and satisfies the estimates
$\ind_G\leq\theta\leq\ind_F$ and
$|\theta(x)-\theta(y)|\leq 3r^{-1}d(x,y)$. 
In particular, a filter $\xi$ is coarse if and only if for any
$F\in\xi$ and any $\varepsilon>0$ there is $G\in\xi$ and
a function $\theta$ such that   $\ind_G\leq\theta\leq\ind_F$ and
$|\theta(x)-\theta(y)|\leq \varepsilon d(x,y)$. 
\end{lemma}
\proof 
If $a\in G$ and $b\notin F$ then $r< d(a,b)\leq d(x,a)+d(x,b)$ for
any $x$. By taking the lower bound of the right hand side over $a,b$
we get  $r\leq d_G(x) + d_{F^\c}(x) \equiv D(x)$. Hence if
$d(x)\equiv d_{F^\c}(x)$ then
\[
|\theta(x)-\theta(y)|\leq \frac{|d(x)- d(y)|}{D(x)} +
d(y) \frac{|D(x)- D(y)|}{D(x)D(y)}\leq \frac{d(x,y)}{r} + 
|D(x)- D(y)|\leq \frac{d(x,y)}{3r}.
\]
To prove the last assertion, notice that if such a $\theta$ exists
for some $\varepsilon<1/r$ and if $x\in G$ and $d(x,y)\leq r$ then
$\theta(x)=1$ and $|\theta(x)-\theta(y)|<1$ hence $\theta(y)>0$ so
$y\in F$. Thus $G_{(r)}\subset F$.  \qed

\begin{proposition}\label{pr:fid}
The filter $\xi$ is coarse if and only if the ideal
$\ci_\xi(X)$ is coarse. 
\end{proposition}
\proof Assume $\xi$ is not trivial and coarse and let
$\varphi\in\ci_\xi$ positive and $r>0$. Then
$\co_\varphi:=\{\varphi<1\}\in\xi$ hence there is $G\in\xi$
such that $G_{(2r)}\subset\co_\varphi$. By using Lemma \ref{lm:dfg}
we construct $\psi\in\cc$ such that $0\leq\psi\leq1$, $\psi|_G=0$,
and $\psi|_{G_{(r)}^\c}=1$. Clearly $\psi\in\ci_\xi$. If
$\psi(y)<1$ then $y\in G_{(r)}$ hence if $d(x,y)\leq r$ then $x\in
G_{(2r)}$ so $\varphi(x)<1$.  Thus $\ci_\xi$ is
coarse. Reciprocally, assume that $\ci_\xi$ is a coarse
ideal and let $F\in\xi$ and $r>0$. There is
$\varphi\in\ci_\xi$ positive such that $\co_\varphi\subset
F$ and there is a positive function $\psi\in\ci_\xi$ such
that \eqref{eq:fid} holds. But then $\co_\psi\in\xi$ and
$(\co_\psi)_{(r)}\subset\co_\varphi$ so $\xi$ is coarse.  \qed

\subsection{Coarse envelope} \label{s:env}

If $\xi$ is a filter then the family of coarse filters included in
$\xi$ is admissible, hence there is a largest coarse filter included
in $\xi$. We denote it $\coa(\xi)$ and call it \emph{coarse envelope
  (or cover) of $\xi$}. A set $F$ belongs to $\coa(\xi)$ if and only
if $F^{(r)}\in\xi$ for any $r>0$ (the set of such $F$ is a filter,
see page \pageref{p:fg}).

By Lemma \ref{lm:coarsefr} we have only two possibilities: either
$\coa(\xi)=\{X\}$ or $\coa(\xi)\supset\infty$. Since
$\coa(\xi)\subset\xi$, we see that either $\xi$ is finer
than Fr\'echet, and then  $\coa(\xi)\supset\infty$, or not, and
then $\coa(\xi)=\{X\}$. 

To each ultrafilter $\vkappa\in\beta(X)$ we associate a compact
subset $\widehat\vkappa\subset\beta(X)$ by the rule
\begin{equation}\label{eq:cok}
  \widehat\vkappa:=\coa(\vkappa)^\dagger =\text{ set of ultrafilters
    finer than the coarse envelope of } \vkappa.
\end{equation}
Thus we have either $\vkappa\in\delta(X)$ and then
$\widehat\vkappa\subset\delta(X)$, or $\vkappa\nin\delta(X)$ and
then $\widehat\vkappa=\beta(X)$.  On the other hand, we have
$\bigcup_{\vkappa\in\delta(X)}\widehat\vkappa=\delta(X)$ because
$\vkappa\in\widehat\vkappa$.

More explicitly, if $\vkappa,\chi\in\delta(X)$ then
$\chi\in\widehat\vkappa$ means: if $F$ is a set such that
$F^{(r)}\in\vkappa$ for all $r$, then $F\in\chi$ (which is
equivalent to $F\cap G\neq\emptyset$ for all $G\in\chi$).

If $\vkappa$ is an ultrafilter on $X$ then $\cc_{(\vkappa)}(X)$ is
the coarse ideal of $\cc(X)$ defined by
\begin{equation}\label{eq:cid}
\cc_{(\vkappa)}(X) = \ci_{\coa(\vkappa)}=
\{ \varphi\in\cc(X) \mid {\textstyle\lim_{\coa(\vkappa)}} \varphi=0\}. 
\end{equation}
The quotient $C^*$-algebra $\cc_{\vkappa}(X) = \cc(X)/
\cc_{(\vkappa)}(X)$ will be called \emph{localization of $\cc(X)$ at
  $\vkappa$}. If $\varphi\in\cc(X)$ then its image in the quotient
is denoted $\vkappa.\varphi$ and is called \emph{localization of
  $\vphi$ at $\vkappa$}.  The next comments give another description
of these objects and will make clear that \emph{localization means
  extension followed by restriction}.

Observe that $\varphi\in\cc(X)$ belongs to $\cc_{(\vkappa)}(X)$ if
and only if the restriction of $\beta(\varphi)$ to $\widehat\vkappa$
is zero. Hence two bounded uniformly continuous functions are equal
modulo $\cc_{(\vkappa)}(X)$ if and only if their restrictions to
$\widehat\vkappa$ are equal.  Thus
$\varphi\mapsto\beta(\varphi)|\widehat\vkappa$ \emph{induces an
  embedding $\cc_\vkappa(X)\hookrightarrow C(\widehat\vkappa)$ which
  allows us to identify $\cc_\vkappa(X)$ with an algebra of
  continuous functions on $\widehat\vkappa$}.  From this we deduce
\begin{equation}\label{eq:uint}
\ccap_{\vkappa\in\delta(X)}\cc_{(\vkappa)}(X)=\Co(X).
\end{equation}
Indeed, $\varphi$ belongs to the left hand side if and only if
$\beta(\varphi)|\widehat\vkappa=0$ for all
$\vkappa\in\delta(X)$. But the union of the sets $\widehat\vkappa$
is equal to $\delta(X)$ hence this means
$\beta(\varphi)|\delta(X)=0$ which is equivalent to
$\varphi\in\Co(X)$.

A \emph{maximal coarse filter} is a coarse filter which is maximal
in the set of coarse filters equipped with inclusion as order
relation.  This set is inductive (the union of an increasing set of
coarse filters is a coarse filter) hence each coarse filter is
majorated by a maximal one. Dually, we say that \emph{a subset
  $T\subset\delta(X)$ is coarse} if it is of the form
$T=\vkappa^\dagger$ for some coarse filter $\vkappa$. Note that if
$T$ is a minimal coarse set then $T=\widehat\vkappa$ for any
ultrafilter $\vkappa\in T$.  In general the coarse sets of the form
$\widehat\vkappa$ with $\vkappa\in\delta(X)$ are not minimal.

\section{Ideals of $\re(X)$} \label{s:ide}

There are two classes of ideals in $\re(X)$ which can be defined in
terms of the behavior at infinity of the operators.  For any filter
$\xi$ on $X$ we define
\begin{align}
\rj_{\xi}(X) & =\{T\in\re(X) \mid 
\inf_{F\in\xi}\|\ind_F T\|=0\}, \label{eq:ek1} \\ 
\rg_{\xi}(X) & =\{T\in\re(X) \mid \lim_{x\to\xi}\|\ind_{B_x(r)}T\|=0\
\forall r \}. \label{eq:gxi}
\end{align} 
Here $\inf_{F\in\xi}\|\ind_F T\|$ is the lower bound of the numbers
$\|\ind_F T\|$ when $F$ runs over the set of measurable $F\in\xi$
and we define $\inf_{F\in\xi}\|T\ind_F\|$ similarly. Note that
$\|\ind_F T\|\leq \|\ind_G T\|$ and $\|T\ind_F\|\leq \|T\ind_G\|$ if
$F\subset G$ are measurable.  Recall also that
$\lim_{x\to\xi}\|\ind_{B_x(r)} T\|=0$ means: for each
$\varepsilon>0$ there is $G\in\xi$ such that $\|\ind_{B_x(r)}
T\|<\varepsilon$ for all $x\in G$.
Observe that for the Fr\'echet filter $\xi=\infty$ we have
\begin{equation}\label{eq:jgf}
\rk=\rj_{\infty} \quad\text{and}\quad 
\rk\subset\rg_\infty=\rg
\end{equation}
where $\rg(X)$ is the ghost ideal introduced in \eqref{eq:ghost}.
That $\rj_{\infty}=\rk$ follows from the fact that $\ind_KT$ is
compact if $K$ is compact (or use \eqref{eq:ek3} and Proposition
\ref{pr:kx}). The equality $\rg_\infty(X)=\rg(X)$ is just a change
of notation

\begin{lemma}\label{lm:inf}
  If $T\in\re$ and $\xi$ is a coarse filter then
  $\inf_{F\in\xi}\|\ind_F T\|=\inf_{F\in\xi}\|T\ind_F \|$.
\end{lemma}
\proof If $\inf_{F\in\xi}\|\ind_F T\|=a$ and $\varepsilon>0$ then
there is $F\in\xi$ such that $\|\ind_F T\|<a + \varepsilon$. We may
choose $k\in\ccfr$ such that $\|T-Op(k)\|<\varepsilon$ and then
$\|\ind_F Op(k)\|<a +2\varepsilon$. Assume that $k(x,y)=0$ if
$d(x,y)\geq r$ and let $G\in\xi$ such that $G_{(r)}\subset F$.  Then
$k(x,y)\ind_G(y)=\ind_{G_{(r)}}(x) k(x,y)\ind_G(y)$ hence $
Op(k)\ind_G =\ind_{G_{(r)}} Op(k)\ind_G = \ind_{G_{(r)}} \ind_F
Op(k)\ind_G $ so $\|Op(k)\ind_G \|\leq\|\ind_F Op(k)\|<a +
2\varepsilon$ and so $\|T\ind_G \|<a+3\varepsilon$.  \qed

\begin{lemma}\label{lm:gxi}
  For any filter $\xi$ the set $\rg_\xi$ is an ideal of $\re$ and we
  have $\rj_{\coa(\xi)}\subset\rg_\xi$. If $\xi$ is coarse then
  $\rj_\xi$ is also an ideal of $\re$ and $\rj_{\xi}\subset\rg_\xi$.
\end{lemma}
\proof $\rg_\xi$ is obviously a closed right ideal in $\re$ so it
will be an ideal if show that $\lim_{x\to\xi}\|T\ind_{B_x(r)}\|=0$
for all $T\in\rg_\xi$. Choose $\varepsilon >$ and let $S$ be a
controlled operator such that $\|S-T\|<\varepsilon$. Then there is
$R$ such that $S\ind_{B_x(r)}=\ind_{B_x(R)} S \ind_{B_x(r)}$ and
there is $F\in\xi$ such that $\|\ind_{B_x(R)}T\|<\varepsilon $ for
$x\in F$, hence
$$
\|T\ind_{B_x(r)}\|< \varepsilon+\|S\ind_{B_x(r)}\|
\leq \varepsilon+\|\ind_{B_x(R)}S\| 
< 2\varepsilon+\|\ind_{B_x(R)}T\| < 3\varepsilon. 
$$
If $T\in\rj_{\coa(\xi)}$ then for any $\varepsilon>0$ there is $F$
such that $F^{(r)}\in\xi$ for all $r$ such that $\|\ind_F
T\|<\varepsilon $. So if we fix $r$ and take $G=F^{(r)}\in\xi$ then
$G\in\xi$ and $\|\ind_{B_x(r)}T\| < \varepsilon$ for all $x\in
G$. Thus $T\in\rg_\xi$.  Clearly $\rj_\xi$ is a closed right ideal
in $\re$. That it is an ideal if $\xi$ is coarse follows from Lemma
\ref{lm:inf}.  \qed

\begin{proposition}\label{pr:ek}
If $\xi$ is a coarse filter on $X$ then $\rj_\xi$ is an
ideal of $\re$ and we have
\begin{equation}\label{eq:ek3}
  \rj_\xi=\ci_\xi\re= \re\ci_\xi.
\end{equation}
\end{proposition}
\proof 
We now prove the first equality in \eqref{eq:ek3} (the second
one follows by taking adjoints). Clearly
$\varphi\in\ci_\xi$ if and only if for each $\varepsilon >0$
there is $F\in\xi$ such that $\|\ind_F \varphi \|<\varepsilon$
hence if and only if $\inf_{F\in\xi}\|\ind_F \varphi \|=0$. This
implies $\ci_\xi\re\subset\rj_\xi$ and so it remains to be
shown that for each $T\in\rj_\xi$ there are
$\varphi\in\ci_\xi$ and $S\in\re$ such that $T=\varphi S$. If
$\xi$ is trivial this is clear, so we may suppose
that $\xi$ is finer than $\infty$.

Choose a point $o\in X$ and let $K_n=B_o(n)$ for $n\geq1$
integer. We get an increasing sequence of compact sets such that
$\cup_n K_n=X$ and $K_n^\c\in\xi$. We construct by induction a
sequence $F_1\supset G_1\supset F_2\supset G_2 \dots$ of sets in
$\xi$ such that: 
\[
F_n\subset K_n^\c, \quad
\|\ind_{F_n} T\|\leq n^{-2}, \quad
d(G_n,F_n^\c)>1, \quad
d(F_{n+1},G_n^\c)>1.
\]
We start with $F'_1\in\xi$ such that $\|\ind_{F'_1} T\|\leq1$,
we set $F_{1}=F'_{1}\cap K_{1}^\c$ and then we choose
$G_1\in\xi$ such that $d(G_1,F_1^\c)>1$. Next, we choose
$F_2'\in\xi$ with $\|\ind_{F'_2} T\|\leq1/4$ and
$G_1'\in\xi$ with $G_1'\subset G_1$ and $d(G_1',G_1^\c)>1$. We
take $F_2=F_2'\cap G_1'\cap K_2^\c$, so $d(F_2,G_1^\c)>1$, and then
we choose $G_2\in\xi$ with $G_2\subset F_2$ such that
$d(G_2,F_2^\c)>1$, and so on.

Now we use Lemma \ref{lm:dfg} and for each $n$ we construct a
function $\theta_n\in\cc$ such that
$\ind_{G_n}\leq\theta_n\leq\ind_{F_n}$ and
$|\theta_n(x)-\theta_n(y)|\leq 3d(x,y)$. Then either $B_a\cap
F_1=\emptyset$ or there is a unique $m$ such that $B_a\cap
F_m\neq\emptyset$ and $B_a\cap F_{m+1}=\emptyset$ and in this case
$\theta_n=1$ on $B_a$ if $n<m$ and $\theta_n=0$ on $B_a$ if
$n>m$. Let $\theta(x)=\sum_n\theta_n(x)$. Then $\theta(x)=0$ on
$F_1^\c$ and if $B_a\cap F_m\neq\emptyset$ and $B_a\cap
F_{m+1}=\emptyset$ we get
\begin{equation}\label{eq:m}
\theta(x) =\sum_{n\leq m}\theta_n(x) =m-1+\theta_m(x).
\end{equation}
Thus $\theta:X\to\bar{\R}_+$ is well defined and for $d(x,y)<1$ and a
conveniently chosen $m$  we have
\[
|\theta(x)-\theta(y)|=|\theta_m(x)-\theta_m(y)|\leq3d(x,y).
\]
On the other hand $\|\theta_n T\| \leq \|\ind_{F_n} T\|\leq n^{-2}$.
Thus if $\theta_0=1$ then the limit of $\sum_{n\leq m}\theta_n T$ as
$m\to\infty$ exists in norm and defines an element $S$ of
$\re$. Then
\[
T=\big(\textstyle{\sum_{n\leq m}}\theta_n  \big)^{-1}
\big(\textstyle{\sum_{n\leq m}}\theta_n  \big)T \to 
(1+\theta )^{-1} S
\]  
because 
$\big(\textstyle{\sum_{n\leq m}}\theta_n \big)^{-1}
\to(1+\theta )^{-1}$ strongly on $L^2(X)$. If
$\varphi:=(1+\theta)^{-1}$ then $0\leq\varphi\leq1$ and 
\[
|\varphi(x) - \varphi(y)| \leq
|\theta(x) - \theta(y)|\leq 3 d(x,y) \quad\text{if } d(x,y)<1.
\]
Thus $\varphi\in\cc$. If $x\in B_a$ with
$B_a\cap F_m\neq\emptyset$ and $B_a\cap F_{m+1}=\emptyset$ then
\eqref{eq:m} gives
\[
\varphi(x)=(1+m-1+\theta_m(x))^{-1}\leq 1/m
\]
hence $\varphi(x)\leq 1/m$ on $F_m$. Thus $\lim_\xi\varphi=0$
and $T=\varphi S$ with $\varphi\in\ci_\vkappa$ and $S\in\re$.
\qed

We make now more precise the relation between $\rj_\xi$ and
$\rg_\xi$.

\begin{lemma}\label{lm:xi}
  If \eqref{eq:12} holds, $T\in\re$ is controlled, $\xi$ is coarse,
  and $\lim_{x\to\xi}\|\ind_{B_x}T\|=0$, then $T\in\rj_\xi$.
\end{lemma}
\proof Assume \eqref{eq:12} is satisfied and let $T\in\rb(X)$ be a
controlled operator. Let $Z$ be as in Lemma \ref{lm:capacity} and
let us set $a=d(T)+1$, so that
$\ind_{B_x}T=\ind_{B_x}T\ind_{B_x(a)}$ for all $x$. If $F$ is a
measurable set and if we denote $Z(F)$ the set of $z\in Z$ such that
$B_z\cap F\neq\emptyset$ then for any $f\in L^2(X)$ we have
\begin{align*}
\|\ind_F Tf\|^2 & \leq \sum_{z\in Z(F)} \|\ind_{B_z} T f\|^2 =
\sum_{z\in Z(F)} \|\ind_{B_z} T\ind_{B_z(a)} f\|^2 \\
&\leq \sup_{z\in Z(F)}\|\ind_{B_z} T\|^2
\sum_{z\in Z(F)} \|\ind_{B_z(a)} f\|^2
\leq \sup_{x\in F_{(1)}}\|\ind_{B_x} T\|^2 N(a) \|f\|^2
\end{align*}
so $\|\ind_F T\|\leq N(a)^{1/2} \sup_{x\in F_{(1)}}
\|\ind_{B_x}T\|$. Thus for any controlled operator we have
$\inf_{F\in\xi}\|\ind_F T \|=0$ if $\lim_{x\to\xi}\|\ind_{B_x}T
\|=0$. If $T\in\re(X)$ this means $T\in\rj_\xi$. \qed

\begin{proposition}\label{pr:xi}
  If $X$ is a class A space then for any filter $\xi$ finer than
  Fr\'echet we have $\rj_{\coa(\xi)}\subset\rg_\xi$. If $\xi$ is
  coarse and $T\in\re$ then
\begin{equation}\label{eq:xib}
T\in\rj_\xi \Leftrightarrow 
\lim_{x\to\xi}\|T\ind_{B_x}\|=0 \Leftrightarrow 
\lim_{x\to\xi}\|\ind_{B_x}T\|=0.
\end{equation} 
\end{proposition}
\proof We use the same techniques as in the proof of Theorem
\ref{th:gc}. Let $T\in\re(X)$ and let us assume that
$\lim_{x\to\xi}\|T\ind_{B_x}\|=0$. Then as we saw in Section
\ref{s:ell} we have $(T\ind_{B_x})_\phi=T_\phi\ind_{B_x}$ hence for
conveniently chosen $\phi$ the operator $T_\phi\in\re(X)$ is
controlled and $\lim_{x\to\xi}\|T_\phi\ind_{B_x}\|=0$. From Lemma
\ref{lm:xi} we get $T_\phi\in\rj_\xi(X)$ which is closed, so since
$T_\phi \to T$ in norm as $\phi\to 1$, we get $T\in\rj_\xi(X)$. \qed

\begin{remark}\label{re:xi}{\rm The relation \eqref{eq:xib} is not
    true in general if Property A is not satisfied. Indeed, if we
    take $\xi=\infty$ then this would mean $\rk=\rg$, which
    does not hold in general.  }\end{remark}

We now seek for a more convenient description of
$\rj_{\coa(\xi)}$ for not coarse filters.

\begin{remark}\label{re:sets}{\rm The following observations are
    easy to prove and will be useful below. Let $F$ be any subset of
    $X$ and let $r,s>0$. Then  
    $F^{(r+s)} \subset (F^{(r)})^{(s)} $ and if $0<r<s$ then
    $F^{(s)} \subset F^{(r)}$ and $F \subset (F_{(s)})^{(r)} $.
%    and $(F^{(s)})_{(r)} \subset F^{(s-r)}$.  
  }\end{remark}

\begin{proposition}\label{pr:bxr}
  Assume that \eqref{eq:12} is satisfied and let $T$ be a controlled
  operator and $\xi$ a filter finer than the Fr\'echet filter. Then
  $\inf_{F\in\coa(\xi)}\|\ind_F T\|=0$ if and only if
  $\lim_{x\to\xi}\|\ind_{B_x(r)}T\|=0$ for all $r>0$.
\end{proposition}
\proof If $T\in\rb(X)$ and $\inf_{F\in\coa(\xi)}\|\ind_F T\|=0$ then
the first few lines of the proof of Lemma \ref{lm:xi} give
$\lim_{x\to\coa(\xi)}\|\ind_{B_x(r)}T\|=0$ for all $r>0$, which is
more than required. Now let $T$ be a controlled operator and let us
set $a=d(T)+1$. If $F$ is a measurable set and $Z(F)$ is as in the
proof of Lemma \ref{lm:xi} then $d(F,Z(F)\leq 1$ hence for any $r>0$
we have
$$
F_{(r)}\subset Z(F)_{(r+1)}=\ccup_{z\in Z(F)} B_z(r+1)
$$
hence for any $f\in L^2$ we have
\begin{align*}
\|\ind_{F_{(r)}}  Tf\|^2 & \leq \sum_{z\in Z(F)} \|\ind_{B_z(r+1)} T f\|^2 =
\sum_{z\in Z(F)} \|\ind_{B_z(r+1)} T\ind_{B_z(r+a)} f\|^2 \\
&\leq \sup_{z\in Z(F)}\|\ind_{B_z(r+1)} T\|^2
\sum_{z\in Z(F)} \|\ind_{B_z(r+a)} f\|^2
\leq \sup_{x\in F_{(1)}}\|\ind_{B_x(r+1)} T\|^2 N(r+a) \|f\|^2.
\end{align*}
If $x\in F_{(1)}$ and $y\in F$ is such that $d(x,y)\leq 1$ then
$B_x(r+1)\subset B_y(r+2)$ hence we obtain
\begin{equation}\label{eq:ndt}
\|\ind_{F_{(r)}}  T\| \leq N(r+a)^{1/2} 
\sup_{x\in F}\|\ind_{B_x(r+2)} T\|
\end{equation}
Observe also that for an arbitrary measurable set $G$ we have the
estimate
\begin{equation}\label{eq:estim}
  \|\ind_G T\|\leq N(a)^{1/2}\sup_{x\in X}   \|\ind_{G\cap B_x} T\|.
\end{equation}
This follows from  Lemma \ref{lm:estime} after noticing that
$d(\ind_G T)\leq d(T)$.  

Now assume that $\lim_{x\to\xi}\|\ind_{B_x(r)}T\|=0$ for all $r>0$
and let us fix $\varepsilon>o$. Then for each $r>0$ there is
$F^r\in\xi$ such that
$$
\|\ind_{B_x(r+2)}T\|\leq \varepsilon N(r+a)^{-1/2} N(a)^{-1/2}
\quad \forall x\in F^r.
$$
For each $f\in L^2$ and each number $s>0$ the map
$x\mapsto\ind_{B_x(s)}f\in L^2$ is strongly continuous, hence the
function $x\mapsto\|\ind_{B_x(r+2)}T\|$ is lower semi-continuous, so
we may assume that $F^r$ is closed, hence measurable. Then the
$G_r:=F^r_{(r)}\in\xi$ is closed and $\|\ind_{G_r} T\|\leq
\varepsilon N(a)^{-1/2}$ because of \eqref{eq:ndt}. Moreover, if
$\alpha<r$ then $G_r^{(\alpha)}\equiv(G_r)^{(\alpha)} \supset F^r$
hence $G_r^{(\alpha)}\in\xi$. Now fix $\alpha>1$ and let
$G=\cup_{r>\alpha} G_r^{(\alpha)}$. This is a union of open set
hence it is open and contains all the $G_r^{(\alpha)}$, which belong
to $\xi$, hence belongs to $\xi$. If $s>0$ and we choose some
$r>s+\alpha$ then $ G^{(s)}\supset (G_r^{(\alpha)})^{(s)} \supset
G_r^{(\alpha+s)} \in \xi$ (Remark \ref{re:sets}). Thus we see that
$G^{(s)}\in\xi$ for all $s>0$, which means that $G\in\coa(\xi)$. In
order to estimate the norm of $\ind_G T$ we use \eqref{eq:estim} and
observe that if $G\cap B_x\neq\emptyset$ the there is $r>\alpha$
such that $G_r^{(\alpha)}\cap B_x\neq\emptyset$ hence $B_x\subset
(G_r^{(\alpha)})_{(1)}$. But it is easy to check that
$(G_r^{(\alpha)})_{(1)}\subset G_r$ because $\alpha>1$, hence
$B_x\subset G_r$, and then
$$
\|\ind_{G\cap B_x} T\|\leq \|\ind_{B_x} T\| \leq
\|\ind_{G_r} T\| \leq \varepsilon N(a)^{-1/2}.
$$
Finally, from \eqref{eq:estim} we get $\|\ind_{G}
T\|\leq\varepsilon$.
\qed

\begin{theorem}\label{th:nyx}
  Let $X$ be a class A space and let $\xi$ be a filter finer than
  Fr\'echet on $X$. If $T\in\re$ then:
\begin{equation}\label{eq:nyx}
T\in\rj_{\coa(\xi)} \Leftrightarrow 
\lim_{x\to\xi}\|T\ind_{B_x(r)}\|=0 \ \forall r>0 \Leftrightarrow 
\lim_{x\to\xi}\|\ind_{B_x(r)}T\|=0 \ \forall r>0.
\end{equation} 
\end{theorem}
\proof This is a repetition of the proof of Proposition
\ref{pr:xi}. For example, let $\lim_{x\to\xi}\|\ind_{B_x(r)}T\|=0$
for all $r>0$. Since $(\ind_{B_x(r)} T)_\phi=\ind_{B_x(r)}T_\phi$
for all $r$, we see that for conveniently chosen $\phi$ the operator
$T_\phi\in\re(X)$ is controlled and
$\lim_{x\to\xi}\|T_\phi\ind_{B_x(r)}\|=0$ for all $r$. From
Proposition \ref{pr:bxr} we clearly get $T_\phi\in\rj_{\coa(\xi)}$
which is closed. So $T\in\rj_{\coa(\xi)}$ because $T_\phi \to T$ in
norm as $\phi\to 1$. \qed

The ideals of $\re(X)$ which are of real interest in our context are
defined as follows
\begin{equation}\label{eq:ideals}
\vkappa\in\delta(X) \Rightarrow
\re_{(\vkappa)}(X):=\rj_{\coa(\vkappa)}(X)
=\{T\in\re(X) \mid 
\inf_{F\in\coa(\vkappa)}\|\ind_F T\|=0\}.
\end{equation}
By Proposition \ref{pr:ek} this can be expressed in terms of the
ideals of $\cc(X)$ introduced in \eqref{eq:cid} as follows:
\begin{equation}\label{eq:ideals1}
\re_{(\vkappa)}(X)=\cc_{(\vkappa)}(X)\re(X)= \re(X)\cc_{(\vkappa)}(X).
\end{equation}

\noindent{\bf Prof of Theorem \ref{th:detailed}: }
Assume that $T\in\re_{(\vkappa)}$ for all $\vkappa\in\delta(X)$; we
have to show that $T$ is a compact operator (the converse being
obvious). If $\vkappa\in\delta(X)$ and $r>0$ then for any
$\varepsilon>0$ there is $F\in\coa(\vkappa)$ such that $\|\ind_F
T\|<\varepsilon $ and there is $G\in\vkappa$ such that
$G_{(r)}\subset F$, hence for any $x\in G$ we have $\|\ind_{B_x(r)}
T\|<\varepsilon$. This proves that
$\lim_{x\to\vkappa}\|\ind_{B_x(r)} T\|=0$. Now define
$\theta(x)=\|\ind_{B_x(r)} T\|$, we obtain a bounded function on $X$
such that $\lim_\vkappa\theta=0$ for any $\vkappa\in\delta(X)$. The
continuous extension $\beta(\theta):\beta(X)\to \R$ has the property
$\beta(\theta)(\vkappa)=\lim_\vkappa\theta$ thus $\beta(\theta)$ is
zero on the compact subset $\delta(X)=\infty^\dagger$ of $\beta(X)$
hence we have $\lim_\infty\theta=0$ according to a remark from
Section \ref{s:filt}.  Thus we have
$\lim_{x\to\infty}\|\ind_{B_x(r)} T\|=0$, which means that $T$
belongs to the ghost ideal $\rg$. Now the compactness of $T$ follows
from Theorem \ref{th:gc}.  \qed

We end this section with some remarks on the case of discrete spaces
with bounded geometry. Assume that $X$ is an infinite set equipped
with a metric $d$ such that the number of points in a ball is
bounded by a number independent of the center of the ball. We equip
$X$ with the counting measure, so $L^2(X)=\ell_2(X)$, and embed
$X\subset\ell_2(X)$ by identifying $x=\ind_{\{x\}}\equiv\ind_{x}$,
so $X$ becomes the canonical orthonormal basis of $\ell_2(X)$. Then
any operator $T\in\rb(X)$ has a kernel $k_T(x,y)=\braket{x}{Ty}$ and
$\re(X)$ is the closure of set of $T$ such that $\braket{x}{Ty} =0$
if $d(x,y)> r(T)$ (this is the uniform Roe algebra). Observe that
for each $T\in\re$ and each $\varepsilon>0$ there is an $r$ such that
$|\braket{x}{Ty}|<\varepsilon$ if $d(x,y)> r$.

If $\xi$ is a filter on $X$ and $f:X^2\to\C$ we write
$\lim_{x,y\to\xi}f(x,y)=0$ if for each $\varepsilon >0$ there is
$F\in\xi$ such that $|f(x,y)|<\varepsilon$ if $x,y\in F$.

\begin{proposition}\label{pr:discrete}
  Let $X$ be discrete with bounded geometry. Then if $\xi$ is a
  filter and $T\in\re$ we have
\begin{equation}\label{eq:discrete1}
  T\in\rg_\xi \Leftrightarrow \lim_{x\to\xi}\sup_{y,z\in B_x(r)}|\braket{y}{Tz}|=0 \quad
  \forall r>0.
\end{equation}
Moreover, if $\xi$ is coarse then
\begin{equation}\label{eq:discrete2}
\rg_\xi =\{ T\in \re \mid \lim_{x,y\to\xi} \braket{x}{Ty}=0 \}.
\end{equation}
\end{proposition}
\proof By definition, we have $T\in\rg_\xi$ if and only if
$\lim_{x\to\xi}\|T\ind_{B_x(r)}\|=0$ for all $r$. Since the norm of
the operator $T\ind_y$ is equal to the norm of the vector $Ty$, we
have
$$
\sup_{y\in B_x(r)} \|Ty\| \leq \|T\ind_{B_x(r)}\| \leq
\sum_{y\in B_x(r)}  \|Ty\| \leq V(r) \sup_{y\in B_x(r)} \|Ty\|.
$$
Thus $T\in\rg_\xi$ is equivalent to $\lim_{x\to\xi}\sup_{y\in
  B_x(r)} \|Ty\|=0$ for all $r$, in particular the property from the
right hand side of \eqref{eq:discrete1} is satisfied. Conversely,
let $T\in\re$ satisfying this condition and let $\varepsilon>0$.
Choose an operator $S$ such that $\|S-T\|<\varepsilon$ and such that
$\braket{x}{Sy}=0$ if $d(x,y)> R$ for some fixed $R$. Then we have $
|\braket{Sy}{a}| \leq \sum_z |\braket{Sy}{z}| |\braket{z}{a}|
\leq\|S\| \sum_{z\in B_y(R)} |\braket{z}{a}| $ hence
\begin{align*}
  \|Ty\|^2 &= \braket{y}{T^*Ty}\leq \varepsilon \|T\| +
  |\braket{Sy}{Ty}| \leq
  \varepsilon \|T\| + \|S\| \sum_{z\in B_y(R)}  |\braket{z}{Ty}| \\
  & \leq \varepsilon \|T\| + \|S\| V(R)\sup_{z\in B_y(R)}
  |\braket{z}{Ty}|
\end{align*}
So for each $\varepsilon>0$ there are $C,R<\infty$ with $\|Ty\|^2
\leq \varepsilon \|T\|+ C\sup_{z\in B_y(R)} |\braket{z}{Ty}|$ for
all $y$.  Hence:
\begin{align*}
\sup_{y\in B_x(r)} \|Ty\|^2  & \leq
\varepsilon \|T\|+ 
C \{ |\braket{z}{Ty}| \mid y\in B_x(r), \ z\in B_y(R) \}
\\
& \leq
\varepsilon \|T\|+ C \sup\{|\braket{z}{Ty}| \mid y,z\in B_x(r+R)  \}.
\end{align*}
This proves the converse implication in \eqref{eq:discrete1}.  

Now assume that $\xi$ is coarse. If $T$ is as in the right hand side
of \eqref{eq:discrete2} then for each $\varepsilon>0$ there is
$F\in\xi$ such that $|\braket{y}{Tz}|<\varepsilon$ if $y,z\in F$ and
for each $r$ there is $G\in \xi$ such that $G_{(r)}\subset F$. Then
if $x\in G$ we have $B_x(r)\subset F$ hence $\sup_{y,z\in
  B_x(r)}|\braket{y}{Tz}|\leq\varepsilon $ so $T\in\rg_\xi$ by
\eqref{eq:discrete1}. Reciprocally, let $T\in\rg_\xi$ and let
$\varepsilon,r>0$. By \eqref{eq:discrete1}, there is $F\in\xi$ such
that if $y,z\in B_x(r)$ for some $x\in F$ then
$|\braket{y}{Tz}|\leq\varepsilon $. Let us choose $r$ such that
$|\braket{y}{Tz}|<\varepsilon$ if $d(y,z)> r$ and let $G\in\xi$ such
that $G_{(r)}\subset F$. If $y,z\in G$ then either $d(y,z)> r$ and
then $|\braket{y}{Tz}|<\varepsilon$, or $d(y,z)\leq r$ and then
$|\braket{y}{Tz}|<\varepsilon$ because $y,z\in B_y(r)$ and $y,z\in
G\subset F$. Thus we found $G\in\xi$ such that
$|\braket{y}{Tz}|<\varepsilon$ if $y,z\in G$. \qed

Finally, for the convenience of the reader we summarize the
construction of the ghost projection of Higson, Laforgue, and
Skandalis. Note that $\rg(X)$ is a $C^*$-algebra of operators on
$\ell_2(X)$ independent of the metric of $X$.  Assume that $X$ is a
disjoint union of finite sets $X_n$ with $1\leq n\leq\infty$ such
that the number $v_n^2$ of elements of $X_n$ tends to infinity with
$n$. Then $\ell_2(X)=\oplus_n\ell_2(X_n)$, the vector
$e_n=\sum_{x\in X_n}x/v_n$ is a unit vector in $\ell_2(X_n)$, and
$\pi:=\sum_n \ket{e_n}\bra{e_n}$ is an orthogonal projection in
$\ell_2(X)$ such that $\braket{x}{\pi y}=0$ if $x,y$ belong to
different sets $X_n$ and $\braket{x}{\pi y}=v_n^{-2}$ if $x,y\in
X_n$. Thus $\pi$ is an infinite rank projection and $\pi\in\rg(X)$.
All this is easy, but the choice of the metric is not: see page 348
in \cite{HLS}.

\section{Locally compact groups} \label{s:groups}

\subsection{Crossed products}\label{s:cp}
In this section we assume that $X$ is a locally compact topological
group with neutral element $e$ and $\mu$ is a left Haar measure. We
write $\d\mu(x)=\d x$ and denote $\Delta$ the modular function
defined by $\d(xy)=\Delta(y)\d x$ or $\d x^{-1} =\Delta(x)^{-1} \d
x$ (with slightly formal notations). There are left and right actions
of $X$ on functions $\varphi$ defined on $X$ given by
$(a.\varphi)(x)=\varphi(a^{-1}x)$ and $(\varphi.a)(x)=\varphi(xa)$.

The left and right regular representation of $X$ are defined by
$\lambda_a f= a.f$ and $\rho_a f=\sqrt{\Delta(a)} f.a$ for $f\in
L^2(X)$.  Then $\lambda_a$ and $\rho_a$ are unitary operators on
$L^2(X)$ which induce unitary representation of $X$ on
$L^2(X)$. These representations commute:
$\lambda_a\rho_b=\rho_b\lambda_a$ for all $a,b\in X$. Moreover, for
$\varphi\in L^\infty(X)$ we have $\lambda_a\varphi(Q)
\lambda_a^*=(a.\varphi)(Q)$ and $\rho_a\varphi(Q)
\rho_a^*=(\varphi.a)(Q)$.

The convolution of two functions $f,g$ on $X$ is defined by
$$
(f*g)(x)=\int f(y) g(y^{-1}x) \d y =
\int f(xy^{-1}) \Delta(y)^{-1}g(y) \d y.
$$
For $\psi\in L^1(X)$ let $\lambda_\psi =\int \psi(y) \lambda_y\d
y\in\rb(X)$. Then $\|\lambda_\psi\|\leq\|\psi\|_{L^1}$ and $\psi*g=
\lambda_\psi g$ for $g\in L^2$.

We recall the definition of the $*$-algebra $L^1(X)$: the product is
the convolution product $f*g$ and the involution is given by
$f^*(x)=\Delta(x)^{-1} \bar{f}(x^{-1})$; the factor $\Delta^{-1}$
ensures that $\|f^*\|_{L^1}=\|f\|_{L^1}$.  The enveloping
$C^*$-algebra of $L^1(G)$ is the \emph{group $C^*$-algebra}
$\cc^*(X)$. The norm closure in $\rb(X)$ of the set of operators
$\lambda_\psi$ with $\psi\in L^1(X)$ is the \emph{reduced group
  $C^*$-algebra} $\cc^*_\rmr(X)$. There is a canonical surjective
morphism $\cc^*(X)\to\cc^*_\rmr(X)$ which is injective if and only
if $X$ is amenable. 

\begin{lemma}\label{lm:rreg}
If $T\in\cc^*_\rmr(X)$ then $\rho_a T=T\rho_a \forall a\in X$.
If $X$ is not compact then $\cc^*_\rmr(X)\cap\rk(X)=\{0\}$.
\end{lemma}
\proof The first assertion is clear because $\rho_a\lambda_b=
\lambda_b\rho_a$. If $X$ is not compact, then $\rho_a\to0$ weakly on
$L^2(X)$ hence if $T\in\cc^*_\rmr(X)$ is compact $\|Tf\|=\|T\rho_a
f\|\to0$ hence $\|Tf\|=0$ for all $f\in L^2(X)$.  \qed

In what follows by uniform continuity we mean ``right uniform
continuity'', so $\varphi:X\to\C$ is uniformly continuous if for any
$\varepsilon>0$ there is a neighborhood $V$ of $e$ such that
$xy^{-1}\in V\Rightarrow |\varphi(x)-\varphi(y)|<\varepsilon$ (see
page 60 in \cite{RS}).  Let $\cc(X)$ be the $C^*$-algebra of bounded
uniformly continuous complex functions.  If $\varphi:X\to\C$ is
bounded measurable then $\varphi\in\cc(X)$ if and only if
$\|\lambda_a\varphi(Q)\lambda_a^*-\varphi(Q)\|\to0$ as $a\to e$.

We consider now crossed products of the form $\ca\rtimes X$ where
$\ca\subset\cc(X)$ is a \mbox{$C^*$-subalgebra} stable under (left)
translations (so $a.\phi\in\ca$ if $\phi\in\ca$; only the case
$\ca=\cc(X)$ is of interest later).  We refer to \cite{Wi} for
generalities on crossed products.  The $C^*$-algebra $\ca\rtimes X$
is the enveloping $C^*$-algebra of the Banach $*$-algebra
$L^1(X;\ca)$, where the algebraic operations are defined as follows:
$$
(f*g)(x)=\int f(y) \, y.g(y^{-1}x) \d y, \quad
f^*(x)=\Delta(x)^{-1} \, x.\bar{f}(x^{-1}).
$$
Thus $\cc^*(X)=\C\rtimes X$. If we define
$\Lambda:L^1(X;\ca)\to\rb(X)$ by $\Lambda(\phi)=\int \phi(a)
\lambda_a \d a$ it is easy to check that this is a continuous
\mbox{$*$-morphism} hence it extends uniquely to a morphism
$\ca\rtimes X\to\rb(X)$ for which we keep the same notation
$\Lambda$. A short computation gives for $\phi\in\Cc(X ;\ca)$ and
$f\in L^2(X)$
$$
(\Lambda(\phi)f)(x) = \int \phi(x,xy^{-1}) \Delta(y)^{-1} f(y) \d y
$$
where for an element $\phi\in\Cc(X ;\ca)$ we set
$\phi(x,a)=\phi(a)(x)$. Thus $\Lambda(\phi)$ is an integral operator
with kernel $k(x,y)=\phi(x,xy^{-1}) \Delta(y)^{-1}$ or
$\Lambda(\phi)=Op(k)$ with our previous notation.

The next simple characterization of $\Lambda$ follows from the
density in $\Cc(X ;\ca)$ of the algebraic tensor product
$\ca\otimes_{\mathrm{alg}}\Cc(X)$: \emph{there is a unique morphism
  $\Lambda:\ca\rtimes X \to \rb(X)$ such that
  $\Lambda(\varphi\otimes\psi)=\varphi(Q)\lambda_\psi$ for
  $\varphi\in\ca$ and $\psi\in \Cc(X)$}. Here we take
$\phi=\varphi\otimes\psi$ with $\varphi\in\ca$ and $\psi\in\Cc(X)$,
so $\phi(a)=\varphi\psi(a)$. Note that the kernel of the operator
$\varphi(Q)\lambda_\psi$ is $k(x,y)=\varphi(x)\psi(xy^{-1})
\Delta(y)^{-1}$.

The reduced crossed product $\ca\rtimes_\rmr X$ is a quotient of the
full crossed product $\ca\rtimes X$, the precise definition is of no
interest here. Below we give a description of it which is more
convenient in our setting. As usual, we embed $\ca\subset\rb(X)$ by
identifying $\varphi=\varphi(Q)$ and if $\rM,\rn$ are subspaces of
$\rb(X)$ then $\rM\cdot\rn$ is the closed linear subspace generated
by the operators $MN$ with $M\in\rM$ and $N\in\rn$.

\begin{theorem}\label{th:rcp}
  The kernel of $\Lambda$ is equal to that of $\ca\rtimes
  X\to\ca\rtimes_\rmr X$, hence $\Lambda$ induces a canonical
  embedding $\ca\rtimes_\rmr X\subset\rb(X)$ whose range is
  $\ca\cdot\cc^*_\rmr(X)$. This allows us to identify
  $\ca\rtimes_\rmr X=\ca\cdot\cc^*_\rmr(X)$.
\end{theorem}

We thank Georges Skandalis for showing us that this is an easy
consequence of results from the thesis of Athina Mageira.  Indeed,
it suffices to take $A=\ca$ and $B=\Co(X)$ in \cite[Proposition
1.3.12]{Ma} by taking into account that the multiplier algebra of
$\Co(X)$ is $\Cb(X)$, and then to use $\Co(X)\rtimes X=\rk(X)$
(Takai's theorem, cf. \cite[Example 1.3.4]{Ma}) and the fact that
the multiplier algebra of $\rk(X)$ is $\rb(X)$.

The crossed product of interest here is $\cc(X)\rtimes_\rmr X =
\cc(X)\cdot\cc^*_\rmr(X)$.  Obviously we have
$\rk(X)=\Co(X)\rtimes_\rmr X\subset\cc(X)\rtimes_\rmr X$, the first
equality being a consequence of Takai's theorem but also of the
following trivial argument: if $\varphi,\psi\in\Cc(X)$ then the
kernel $\varphi(x)\psi(xy^{-1}) \Delta(y)^{-1}$ of the operator
$\varphi(Q)\lambda_\psi$ belongs to $\Cc(X^2)$ hence
$\varphi(Q)\lambda_\psi$ is a Hilbert-Schmidt operator.

We recall that the \emph{local topology} on $\cc(X)\rtimes_\rmr X$
(see Definition \ref{df:local} here and \cite[page 447]{GI}) is
defined by the family of seminorms of the form
$\|T\|_\Lambda=\|\ind_\Lambda T\|+\|T\ind_\Lambda\|$ with
$\Lambda\subset X$ compact.

The following is an extension of \cite[Proposition 5.9]{GI} in the
present context (see also pages 30--31 in the preprint version of
\cite{GI0} and \cite{Ro}). Recall that any bounded function
$\varphi:X\to\C$ extends to a continuous function $\beta(\varphi)$
on $\beta(X)$.  If $\vkappa\in\beta(X)$ we define
$\varphi_\vkappa:X\to\C$ by 
\begin{equation}\label{eq:lim}
\varphi_\vkappa(x)=
\beta(x^{-1}\varphi)(\vkappa)= \lim_{a\to\vkappa}\varphi(xa).
\end{equation}

\begin{lemma}\label{lm:lim}
  If $\varphi\in\cc(X)$ then for any $\theta\in\Co(X)$ the set
  $\{\theta \varphi.a \mid a\in X\}$ is relatively compact in
  $\Co(X)$ and the map $a\mapsto\theta\varphi_a\in\Co(X)$ is norm
  continuous. In particular, for any $\vkappa\in\beta(X)$ the limit
  in \eqref{eq:lim} exists locally uniformly in $x$ and we have
  $\varphi_\vkappa\in\cc(X)$.
\end{lemma}
\proof By the Ascoli-Arzela theorem, to show the relative
compactness of the set of functions of the form $\theta\varphi.a$ it
suffices to show that the set is equicontinuous.  For each
$\varepsilon>0$ there is a neighborhood $V$ of $e$ such that
$|\varphi(x)-\varphi(y)|<\varepsilon$ if $xy^{-1}\in V$.  Then
$|\varphi(xa)-\varphi(ya)|<\varepsilon$ for all $a\in X$, which
proves the assertion.  In particular,
$\lim_{a\to\vkappa}\theta\varphi.a$ exists in norm in $\Co(X)$,
hence the limit in \eqref{eq:lim} exists locally uniformly in
$x$. Moreover, we shall have
$|\varphi_\vkappa(x)-\varphi_\vkappa(y)|<\varepsilon$ so
$\varphi_\vkappa$ belongs to $\cc(X)$. Finally, we show that for any
compact set $K$ and any $\varepsilon>0$ there is a neighborhood $V$
of $e$ such that $\sup_K|\varphi(xa)-\varphi(x)|<\varepsilon$ for
all $a\in V$. For this, let $\cu$ be an open cover of $K$ such that
the oscillation of $\varphi$ over any $U\in\cu$ is $<\varepsilon$
and note that there is an neighborhood $V$ of $e$ such that for any
$x\in K$ there is $U\in\cu$ such that $xV\subset U$ (use the
Lebesgue property for the left uniform structure).  \qed

\begin{proposition}\label{pr:lim}
  For each $T\in\cc(X)\rtimes_\rmr X$ and each $a\in X$ we have
  $\tau_a(T) :=\rho_a T\rho_a^*\in\cc(X)\rtimes_\rmr X$ and the map
  $a\mapsto \tau_a(T)$ is locally continuous on $X$ and has locally
  relatively compact range.  For each ultrafilter
  $\vkappa\in\beta(X)$ and each $T\in\cc(X)\rtimes_\rmr X$ the
  limit $\tau_\vkappa(T):=\lim_{a\to\vkappa}\tau_a(T)$ exists in the
  local topology of $\cc(X)\rtimes_\rmr X$.  The so defined map
  $\tau_\vkappa:\cc(X)\rtimes_\rmr X \to \cc(X)\rtimes_\rmr X$ is a
  morphism uniquely determined by the property
  $\tau_\vkappa(\varphi(Q)\lambda_\psi)=\varphi_\vkappa(Q)\lambda_\psi$.
\end{proposition}
\proof If $T=\varphi(Q)\lambda_\psi$ then $\rho_a
T\rho_a^*=(\varphi.a)(Q)\lambda_\psi$ is an element of
$\cc(X)\rtimes_\rmr X$ and so $\tau_a$ is an automorphism of
$\cc(X)\rtimes_\rmr X$. If we take $\psi$ with compact support and
$\Lambda$ is a compact set then
$\lambda_\psi\ind_\Lambda=\ind_K\lambda_\psi\ind_\Lambda$ where
$K=(\supp\psi)\Lambda$ is also compact. Then
$\tau_a(T)\ind_\Lambda=(\varphi.a)(Q)\ind_K\lambda_\psi\ind_\Lambda$
and the map $a\mapsto(\varphi.a)(Q)\ind_K$ is norm continuous,
cf. Lemma \ref{lm:lim}. This implies that $a\mapsto \tau_a(T)$ is
locally continuous on $X$ for any $T$. To show that the range is
relatively compact, it suffices again to consider the case
$T=\varphi(Q)\lambda_\psi$ with $\psi$ with compact support and to
use
$\tau_a(T)\ind_\Lambda=(\varphi.a)(Q)\ind_K\lambda_\psi\ind_\Lambda$
and the relative compactness of the $\{(\varphi.a)(Q)\ind_K\mid a\in
X\}$ established in Lemma \ref{lm:lim}. The other assertions of the
proposition follow easily from these facts. \qed

\subsection{Elliptic $C^*$-algebra} \label{s:cpe} 

Let $X$ be a locally compact non-compact topological group.
Since we do not require that $X$ be metrizable, we have to adapt
some of the notions used in the metric case to this context. Of
course, we could use the more general framework of coarse spaces
\cite{R} to cover both situations, but we think that the case of
metric groups is already sufficiently general. So the reader may
assume that $X$ is equipped with an invariant proper distance
$d$. Our leftist bias in Section \ref{s:cp} forces us to take $d$
right invariant, i.e.\ $d(x,y)=d(xz,yz)$ for all $x,y,z$. If we set
$|x|=d(x,e)$ then we get a function $|\cdot|$ on $X$ such that
$|x^{-1}|=|x|$, $|xy|\leq|x|+|y|$, and $d(x,y)=|xy^{-1}|$. The balls
$B(r)$ defined by relations of the form $|x|\leq r$ are a basis of
compact neighborhoods of $e$, a function on $X$ is $d$-uniformly
continuous if and only if it is right uniformly continuous, etc.

Note that $B_x(r)=B(r)x$ so in the non-metrizable case the role of
the balls $B_x(r)$ is played by the sets $Vx$ with $V$ compact
neighborhoods of $e$. Recall that the range of the modular function
$\Delta$ is a subgroup of the multiplicative group $]0,\infty[$
hence it is either $\{1\}$ or unbounded. Since
$\mu(Vx)=\mu(V)\Delta(x)$ our assumption \eqref{eq:xrv} is satisfied
only if $X$ is unimodular and in this case we have $\mu(Vx)=\mu(V)$
for all $x$. 

We emphasize the importance of the condition that the metric be
proper. Fortunately, it has been proved in \cite{HP} that a locally
compact group is second countable if and only if its topology is
generated by a proper right invariant metric.

For coherence, in the non metrizable case we are forced to say that
a kernel $k:X^2\to\C$ is \emph{controlled} if there is a compact set
$K\subset X$ such that $k(x,y)=0$ if $xy^{-1}\nin K$. The symbol
$d(k)$ should be defined now as the smallest compact set $K$ with
the preceding property.  On the other hand, $k$ is uniformly
continuous if it is right uniformly continuous, i.e.\ if for any
$\varepsilon>0$ there is a neighborhood $V$ of $e$ such that
$|k(ax,by)-k(x,y)|<\varepsilon$ for all $a,b\in V$ and $x,y\in
X$. Then the Schur estimate \eqref{eq:schur} gives $\|Op(k)\|\leq
\sup|k| \sup_a\mu(Ka)$ so only if $X$ is unimodular we have a simple
estimate $\|Op(k)\|\leq \mu(K)\sup|k|$.

To summarize, if $X$ is unimodular then $\ccfr(X^2)$ is well defined
and Lemma \ref{lm:kl} remains valid if we set $V(d(k))=\mu(d(k))$ so
we may \emph{define the elliptic algebra $\re(X)$ as in
  \eqref{eq:ell}}. But in fact, what we get is just a description of
the crossed product $\cc(X)\rtimes_\rmr X$ independent of the group
structure of $X$:

\begin{proposition}\label{pr:elcr}
  If $X$ is unimodular then $\re(X)=\cc(X)\rtimes_\rmr
  X=\cc(X)\cdot\cc^*_\rmr(X)$.
\end{proposition}
\proof From the results presented in Section \ref{s:cp} and the fact
that $\Delta=1$ we get that $\cc(X)\rtimes X$ is the closed linear
space generated by the operators $Op(k)$ with kernels
$k(x,y)=\varphi(x)\psi(xy^{-1})$, where $\varphi\in\cc(X)$ and
$\psi\in\Cc(X)$. Thus $\cc(X)\rtimes X\subset\re(X)$.  To show the
converse, let $k\in\ccfr(X^2)$ and let $\wtilde k(x,y)=k(x,y^{-1}x)$
hence $k(x,y)=\wtilde k(x,xy^{-1})$. If $K=K^{-1}\subset X$ is a
compact set such that $k(x,y)\neq0 \Rightarrow xy^{-1}\in K$ then
$\supp\wtilde k\subset X\times K$. Fix $\varepsilon>0$ and let $V$
be a neighborhood of the origin such that $|\wtilde k(x,y)-\wtilde
k(x,z)| <\varepsilon$ if $yz^{-1}\in V$.  Then let $Z\subset K$ be a
finite set such that $K\subset\cup_{z\in Z} Vz$ and let
$\{\theta_z\}$ be a partition of unity subordinated to this
covering. If $\wtilde l(x,y) = \sum_{z\in Z} \wtilde k(x,z)
\theta_z(y)$ or $\wtilde l= \sum_{z\in Z} \wtilde k(\cdot,z) \otimes
\theta_z$ then
\[
|\wtilde k(x,y)-\wtilde l(x,y)| =
|\sum_{z\in Z} 
(\wtilde k(x,y)- \wtilde k(x,z))\theta_z(y)|\leq
\sum_{z\in Z} |\wtilde k(x,y)- \wtilde k(x,z)|\theta_z(y)
\leq\varepsilon 
\]
because $\supp\theta_z\subset Vz$. Now let us set $l(x,y)=\wtilde
l(x,xy^{-1})=\sum_{z\in Z}\wtilde k(x,z)\theta_z(xy^{-1})$.  If
$l(x,y)\neq0$ then $\theta_z(xy^{-1})\neq0$ for some $z$ hence
$xy^{-1}\in Vz\subset VK$. In this construction we may choose
$V\subset U$ where $U$ is a fixed compact neighborhood of the
origin. Then we will have $l(x,y)\neq0 \Rightarrow xy^{-1}\subset
UK$ which is a compact set independent of $l$ and from
\eqref{eq:shur} we get $\|Op(k)-Op(l)\|\leq C\sup|k-l|\leq
C\varepsilon$ for some constant $C$ independent of
$\varepsilon$. But clearly $Op(l)\in\cc(X)\rtimes_\rmr X$. \qed

Thus if $X$ is a unimodular group then we may apply Proposition
\ref{pr:lim} and get endomorphisms $\tau_\vkappa$ of $\re(X)$
indexed by $\vkappa\in\delta(X)$. These will play an important role
in the next subsection.

We make now some comments on the relation between amenability and
Property A in the case of groups. First, the Property A is much more
general than amenability, cf. the discussion in \cite{NY} for the
case of discrete groups. To show that amenability implies Property A
we choose from the numerous known equivalent descriptions that which
is most convenient in our context \cite[page 128]{Pa}: \emph{$X$ is
  amenable if and only if for any $\varepsilon>0$ and any compact
  subset $K$ of $X$ there is a positive function $\varphi\in\Cc(X)$
  with $\|\varphi\|=1$ such that
  $\|\rho_a\varphi-\varphi\|<\varepsilon$ for all $a\in K$.}  Now
let us set $\phi(x)=\rho_x^*\varphi$, so
$\phi(x)(z)=\Delta(x)^{-1/2}\varphi(zx^{-1})$. We get a strongly
continuous function $\phi:X\to L^2(X)$ such that $\|\phi(x)\|=1$,
$\supp\phi(x)=(\supp\varphi)x$, and
$\|\phi(x)-\phi(y)\|=\|\rho_{xy^{-1}}\varphi-\varphi\|\leq\varepsilon$
if $xy^{-1}\in K$. In the metric case we get a function as in
Definition \ref{df:yu}, so the metric version
of the Property A is satisfied. \label{p:amen}

\subsection{Coarse filters in groups} \label{s:cfg} 

A filter $\xi$ on a locally compact non-compact group $X$ is called
\emph{round} if the sets of the form $VG=\{xy\mid x \in V, y \in
G\}$, where $V$ runs over the set of neighborhoods of $e$ and $G$
over $\xi$, are a basis of $\xi$. And $\xi$ is (left)
\emph{invariant} if $x\in X, F\in\xi \Rightarrow xF\in\xi$.
Naturally, $\xi$ is \emph{coarse} if for any $F\in\xi$ and any
compact set $K\subset X$ there is $G\in\xi$ such that $KG\subset F$.

The simplicity of the next proof owes much to a discussion with
H. Rugh. In our initial argument Proposition \ref{pr:gcoarse} was a
corollary of Proposition \ref{pr:fid}.

\begin{proposition}\label{pr:gcoarse}
A filter is coarse if and only if it is round and invariant.
\end{proposition}
\proof  Note first that $\xi$ is invariant if
and only if for each $H\in\xi$ and each finite $N\subset X$ there is
$G\in\xi$ such that $H\supset NG$. This is clear because $NG\subset
H$ is equivalent to $G\subset\cap_{x\in N}x^{-1}H$. Now assume that
$\xi$ is also round. Then for any $F\in\xi$ there is a neighborhood
$V$ of $e$ and a set $H\in\xi$ such that $F\supset VH$. If $K$ is
any compact set then there is a finite set $N$ such that $VN\supset
K$. Then there is $G\in\xi$ such that
$H\supset NG$. So $F\supset VNG \supset KH$. \qed

\begin{proposition}\label{pr:coarsefi}
  Let $X$ be unimodular and let $\xi$ be a coarse filter. Then for
  any $T\in\rj_\xi(X)$ we have $\lim_{a\to\xi}\tau_a(T)=0$
  locally. If $X$ is amenable then the converse assertion holds, so
\begin{equation}\label{eq:gcideal}
  \rj_\xi(X)=\{T\in\re(X) \mid 
\lim_{a\to\xi}\tau_a(T)=0 \text{ locally}\} =
\{T\in\re(X) \mid \tau_\vkappa(T)=0 \ \forall \vkappa\in\xi^\dagger\}.
\end{equation}
Moreover, if $X$ is amenable then for any compact neighborhood $V$
of $e$ and any $T\in\re(X)$ we have:
\begin{equation}\label{eq:cfamenable}
T\in\rj_\xi(X)
\Leftrightarrow \lim_{a\to\xi}\|T\ind_{Va}\|=0
\Leftrightarrow \lim_{a\to\xi}\|\tau_a(T)\ind_{V}\|=0
\end{equation} 
\end{proposition}
\proof We have $\ind_{Va}(Q)=\rho_a^*\ind_V(Q)\rho_a$ hence
$\|T\ind_{Va}\|=\|T\rho_a^*\ind_V(Q)\rho_a\|=
\|\tau_a(T)\ind_V(Q)\|$ hence for $T\in\rj_\xi(X)$ we have
$\lim_{a\to\xi}\tau_a(T)=0$ locally. If $X$ is amenable then
Proposition \ref{pr:xi} in the metric case and a suitable
modification in the non-metrizable group case gives
\eqref{eq:gcideal}. Then \eqref{eq:cfamenable} is easy. \qed

\begin{theorem}\label{th:trans}
  Let $X$ be a unimodular amenable locally compact group. Then for
  each $\vkappa\in\delta(X)$ and for each $T\in\re(X)$ the limit
  $\tau_\vkappa(T):=\lim_{a\to\vkappa} \rho_a T \rho_a^*$ exists in
  the local topology of $\re(X)$, in particular in the strong
  operator topology of $\rb(X)$. The maps $\tau_\vkappa$ are
  endomorphisms of $\re(X)$ and $\bigcap_{\chi\in\delta(X)}
  \ker\tau_\chi=\rk(X)$. In particular, the map
  $T\mapsto(\tau_\vkappa(T))$ is a morphism
  $\re(X)\to\prod_{\vkappa\in\delta(X)}\re(X)$ with $\rk(X)$ as
  kernel, hence the essential spectrum of any normal operator
  $H\in\re(X)$ or any observable $H$ affiliated to $\re(X)$ is given
  by $\spe(H)=\overline\ccup_\vkappa \sp(\tau_\vkappa(H))$.
\end{theorem}
\proof We have seen in Section \ref{s:env} that
$\bigcup_{\vkappa\in\delta(X)}\widehat\vkappa=\delta(X)$ and from
\eqref{eq:gcideal} we get
\begin{equation}\label{eq:trans}
\re_{(\vkappa)}(X)=\ccap_{\chi\in\what{\vkappa}} \ker\tau_\chi \quad
\text{for each } \vkappa\in\delta(X). 
\end{equation}
On the other hand, we have shown before that
$\cap_{\vkappa\in\delta(X)}\re_{(\vkappa)}(X)=\rk(X)$ is a
consequence of Property A, hence of amenability. \qed

\begin{remark}\label{re:bs}{\rm Recall that after \eqref{eq:local}
    we defined the localization $\vkappa.T$ at $\vkappa\in\delta(X)$
    of some $T\in\re$ as the quotient of $T$ in
    $\re_\vkappa=\re/\re_{(\vkappa)}$. If $T$ is normal then from
    \eqref{eq:trans} we get
    $\sp(\vkappa.T)=\overline\ccup_{\chi\in\what\vkappa}
    \sp(\tau_\chi(T))$ but many of the operators $\tau_\chi(T)$
    which appear here are unitary equivalent, in particular have the
    same spectrum. Indeed, note that there is a natural (left)
    action of $X$ on $\beta(X)$ which leaves $\delta(X)$ invariant
    and $\what\vkappa$ is the minimal closed invariant subset of
    $\delta(X)$ which contains $\vkappa$. And if $\chi\in\delta(X)$
    and $a\in X$ then by using $a\chi=\lim_{b\to\chi} ab$ we get
    $\tau_{a\chi}(T)=\rho_a\tau_\chi(T)\rho_a^*$. }\end{remark}

\section{Quasi-controlled operators}\label{s:psd}

In this section we describe briefly other $C^*$-algebras of
operators which are analogs of $\re(X)$. We emphasize that our
choice of $\re(X)$ was determined by our desire to mimic the crossed
product $\cc(X)\rtimes X$ which is a very natural object in the
abelian group case, but there are of course many other
possibilities. For example, we could allow bounded Borel (instead of
uniformly continuous) kernels in \eqref{eq:frk}. The
\mbox{$C^*$-algebra} generated by such kernels is strictly larger
than $\re$ (even if we require the kernels to be continuous, see
Example \ref{ex:cex}) but an analogue of Theorem \ref{th:detailed}
remains true. It is not clear to us if this algebra is really
significant in applications, the set of observables affiliated to
$\re$ being already very large.

We now consider the $C^*$-algebra obtained as norm closure of the
set of controlled operator. This notion has been introduced in the
metric case in Section \ref{s:ell} but in fact it makes sense in the
general framework of coarse spaces $X$ and geometric Hilbert
$X$-modules \cite{R}. In particular, if $X$ is a locally compact
group an operator $T\in \rb(X)$ is \emph{controlled} if there is a
compact set $\Lambda\subset X$ such that if $F,G$ are closed subsets
of $X$ with $F\cap (\Lambda G)=\emptyset$ then $\ind_F T\ind_G
=0$. If $X$ is a metric group with a metric as in Section
\ref{s:cpe} this is equivalent to the definition of Section
\ref{s:ell}.  We denote $\rc(X)$ the norm closure of the set of
controlled operators and we call \emph{quasi-controlled operators}
its elements. If $X$ is a proper metric space this is the ``standard
algebra" from \cite{D}.  If $X$ is a discrete metric space with
bounded geometry then $\rc(X)=\re(X)$ is the "uniform Roe
$C^*$-algebra" from \cite{R,CW1,CW2,Wa}. Clearly $\rc(X)\supset
\re(X)$.

One may define analogs of the ideals $\rj_\xi$ and
$\rg_\xi$. Indeed, form the proof of Lemma \ref{lm:inf} it follows
that if $\xi$ is a coarse filter on $X$ then the set $\rj_\xi(X)$ of
$T\in\rc(X)$ such that $\inf_{F\in\xi}\|\ind_F T\|=0$ is an ideal of
$\rc(X)$. And if $\xi$ is an arbitrary filter then the set
$\rg_\xi(X)$ of $T\in\rc(X)$ such that
$\lim_{x\to\xi}\|\ind_{\Lambda x} T\|=0$ for each compact set
$\Lambda$ is also an ideal of $\rc(X)$.  But if $X$ is not discrete
this class of ideals is too small to allow one to describe the
quotient $\rc(X)/\rk(X)$ even in simple cases. For example, if
$X=\R$ then the operators in $\rc$ may have an anisotropic behavior
in momentum space (see Proposition \ref{pr:rcxa} and \cite{GIc}).

In order to clarify the difference between $\re(X)$ and $\rc(X)$ we
consider the case when $X$ is an abelian group.  We first recall a
result from \cite{GI}). Let $X^*$ be the dual group and for $p\in
X^*$ let $\nu_p$ be the unitary operator on $L^2(X)$ given by
$(\nu_p f)(x)=p(x)f(x)$. To any Borel function $\psi$ on $X^*$ we
associate an operator $\psi(P)=\cf^{-1}M_\psi\cf$ on $L^2(X)$, where
$M_\psi$ is the operator of multiplication by $\psi$ on $L^2(X^*)$
and $\cf$ is the Fourier transformation.

\begin{proposition}\label{pr:ag}
  If $X$ is an abelian group then $\re(X)=\cc(X)\rtimes
  X=\cc(X)\rtimes_\rmr X$ is the set of operators $T\in \rb(X)$ such
  that $\|\nu_p T \nu_p^*-T\|\to0$ and
  $\|(\lambda_a-1)T^{(*)}\|\to0$ if $p\to e$ in $X^*$ and $a\to e$
  in $X$.
\end{proposition}

The equality $\re(X)=\cc(X)\rtimes X$ has been proved before in a
more general setting. Proposition \ref{pr:ag} gives in fact a
description of the crossed product $\cc(X)\rtimes X$ if $X$ is
abelian. If we accept it, then we get the following easy proof of
the inclusion $\re(X)=\cc(X)\rtimes X$. The operators $\nu_p Op(k)
\nu_p^*$ and $\lambda_a Op(k)$ have kernels
$p(x)k(x,y)\bar{p}(y)=p(xy^{-1})k(x,y)$ and $k(xa^{-1},y)$.  Hence
from \eqref{eq:shur} we get
$$ 
\|\nu_p Op(k) \nu_p^*-Op(k)\|\leq
{\textstyle\sup_{xy^{-1}\in K}} |p(xy^{-1})-1| |k(x,y)| \mu(K) 
$$
which tends to zero as $p\to e$ in $X^*$. Similarly $\|(\lambda_a
-1)Op(k)\|\to0$ as $a\to e$ in $X$.  Hence $Op(k)\in \cc(X)\rtimes
X$ for each $k\in\ccfr(X^2)$.

The next example shows the role played by the uniform continuity
condition in the definition of $\re(X)$.

\begin{example}\label{ex:cex}{\rm 
If $X=\R$ then we identify $X^*=\R$ by setting 
$p(x)=\rme^{\rmi px}$. Then the elliptic algebra can be described in
very simple terms. Indeed, if $\lambda_a,\nu_a$ are the unitary
operators on $L^2(\R)$ given by $(\lambda_af)(x)=f(x-a)$ and
$(\nu_af)(x)=\rme^{\rmi ax}f(x)$, we have 
$$
\re(\R)=\{T\in\rb(\R) \mid \|(\lambda_a-1)T^{(*)}\|\to0 \text{ and }
\|\nu_aT\nu_a^*-T\|\to0 \text{ as } a\to0\}. 
$$
Here $T^{(*)}$ means that the relation holds for $T$ and $T^*$. If
we take $k(x,y)=\varphi(x)\theta(x-y)$ with $\varphi\in\cc(\R)$ and
$\theta\in\Cc(\R)$ then $Op(k)=\varphi(Q)\psi(P)\in\re(\R)$ with
$\psi$ the Fourier transform (conveniently normalized) of
$\theta$. The advantage now is that we can see what happens if
$\varphi$ is only bounded and continuous.  Then it is easy to check
that $\varphi(Q)\psi(P)\in\re(\R)$ if and only if
$\|(\varphi(Q+a)-\varphi(Q))\psi(P)\|\to0$ when $a\to0$. For
example, if $\varphi(x)=\rme^{\rmi x^2}$ the last condition is
equivalent to $\|(\rme^{\rmi aQ}-1)\psi(P)\|\to0$, which is
equivalent to $\psi(P)=\eta(Q)S$ for some $\eta\in\Co(\R)$ and
$S\in\rb(\R)$. But then $\psi(P)$ is compact as a norm limit of
operators of the form $\zeta(Q)\psi(P)$ with $\zeta\in\Co(\R)$,
which is not true if $\psi\neq0$. Thus, \emph{the operator
  associated to a kernel of the form $k(x,y)=\rme^{\rmi
    x^2}\theta(x-y)$ with $\theta\in\Cc^\infty(\R)$ and not zero
  does not belong to $\re(\R)$}.  }\end{example}

To describe $\rc(X)$, we need an analogue of Lemma \ref{lm:capacity}
in the group context.

\begin{lemma}\label{lm:part}
  Let $\omega$ be a compact neighborhood of $e$ and $Z$ a maximal
  $\omega$-separated subset of $X$ (i.e.\ if $a,b$ are distinct
  elements of $Z$ then $(\omega a)\cap (\omega b)=\emptyset$). Then
  for any compact set $K\supset\omega^{-1}\omega$ we have $KZ=X$ and
  for any $a\in Z$ the number of $z\in Z$ such that
  $(Kz)\cap(Ka)\neq\emptyset$ is at most
  $\mu(\omega K^{-1}K)/\mu(\omega)$.
\end{lemma}
\proof That such maximal $Z$ exist follows from Zorn lemma.  By
maximality, $(\omega x)\cap (\omega Z)\neq\emptyset$ for any $x$,
hence $x\in \omega^{-1}\omega Z$, so $X= KZ$ if
$K\supset\omega^{-1}\omega$. Now fix $a\in Z$ and let $N$ be the
number of points $z\in Z$ such that $(Kz)\cap(Ka)\neq\emptyset$. For
each such $z$ we have $z\in K^{-1}Ka$ hence $\omega z\subset \omega
K^{-1}Ka$. But the sets $\omega z$ are pairwise disjoint and have
the same measure $\mu(\omega)$ so
$N\mu(\omega)\leq\mu(\omega K^{-1}Ka)=\mu(\omega K^{-1}K)$. \qed

If $X$ is an abelian group then a \emph{$Q$-regular operator} is an
operator $T\in\rb(X)$ which satisfies only the first condition from
Proposition \ref{pr:ag}, i.e.\ is such that the map $p\mapsto\nu_p T
\nu_p^*$ is norm continuous. These operators form a
\mbox{$C^*$-algebra} which contains $\re(X)$, strictly if $X$ is not
discrete, which seems to depend on the group structure of $X$. But
in fact this is not the case, it depends only on the coarse
structure of $X$.

\begin{proposition}\label{pr:rcxa}
  If $X$ is an abelian group then $\rc(X)=\{T\in\rb(X)\mid\lim_{p\to
    e} \|\nu_p T \nu_p^*-T\|=0\}$.
\end{proposition}

For the proof, it suffices to use \cite[Propositions 4.11 and
4.12]{GG} (arXiv version) and Lemma \ref{lm:part}.

Now let $\rl\rc(X)$ be the set of locally compact operators in
$\rc(X)$. Obviously $\rl\rc$ is a $C^*$-algebra and
$\re\subset\rl\rc\subset\rc$ strictly in general. Indeed, let $X$ be
an abelian group, $\vphi$ a bounded continuous function on $X$, and
$\psi\in\cc(X^*)$. Then $\phi(Q)\psi(P)$ belongs to $\rc$ but not to
$\rl\rc$ in general, and if $\psi\in\Co(X^*)$ it belongs to $\rl\rc$
but not to $\rc$ in general, cf.\ Example \ref{ex:cex}. Note that an
operator $T\in\rc$ is locally compact if and only if $\lim_{a\to
  e}\lambda_a T^{(*)}=T^{(*)}$ in the local topology of $\rc$.

Finally, we mention another $C^*$-algebra which is of a similar
nature to $\rc(X)$ and makes sense and is useful in the context of
arbitrary locally compact spaces $X$ and arbitrary geometric Hilbert
$X$-modules, see \cite{GG,R}.  Let us say that $S\in B(\ch)$ is
\emph{quasilocal} (or "decay preserving") if for each
$\varphi\in\Co(X)$ there are operators $S_1,S_2\in B(\ch)$ and
functions $\varphi_1,\varphi_2\in\Co(X)$ such that
$S\varphi(Q)=\varphi_1(Q)S_1$ and $\varphi(Q)S=S_2\varphi_2(Q)$.
The set of quasilocal operators is a $C^*$-algebra which contains
strictly $\rc(X)$ if $X$ is a locally compact non-compact abelian
group. Indeed, if $\psi\in L^\infty(X^*)$ has compact support then
$\psi(P)$ is quasilocal (because $\psi(P)\varphi(Q)$ and $\varphi(Q)
\psi(P)$ are compact) but it belongs to $\rc(X)$ if and only if
$\psi$ is continuous.

\begin{acknowledgment}
  I am grateful to Hans-Henrik Rugh, Armen Shirikyan and Georges
  Skandalis, several discussions with them were very helpful.
\end{acknowledgment}


\begin{thebibliography}{ABG}

\bibitem[ABG]{ABG} { W.\,Amrein, A.\,Boutet de Monvel,
V.\,Georgescu}, {\em $C_0$-Groups, Commutator Methods and Spectral
Theory of $N$-Body Hamiltonians\/}, Birkh\" auser 1996.

\bibitem[Be1]{Be1} { J.\,Bellissard}, {\em Gap labelling theorems for
  Schr\"odinger operators}, in { From Number Theory to Physics\/
    (Les Houches 1989)}, J.M.\,Luck, P.\,Moussa, M.\,Waldschmidt
  (eds.), pp.\,538--630, Springer 1993.

\bibitem[Be2]{Be2} { J.\,Bellissard}, {\em Non Commutative Methods
    in Semiclassical Analysis\/}, Lecture Notes in Math. {\bf 1589},
  Springer 1994.

\bibitem[BG1]{BoG1} { A.\,Boutet de Monvel, V.\,Georgescu}, {\em
    Graded $C^*$-algebras in the $N$-body problem}, {
      J.\,Math.\,Phys.}  {\bf 32} (1991), 3101--3110.

  \bibitem[BG2]{BoG2} A.\,Boutet de Monvel, V.\,Georgescu, {\em
      Graded $C^*$-algebras associated to symplectic spaces and
      spectral analysis of many channel Hamiltonians}, in { Dynamics
      of Complex and Irregular Systems (Bielefeld 1991)}, Bielefeld
    Encount.  Math.\,Phys., vol.\,8 pp.\,22--66 World
    Sci.\,Publishing, River Edge, NJ, 1993.

  \bibitem[CL]{CL} S.\,N.\,Chandler-Wilde, M.\,Lindner, {\it Limit
      Operators, Collective Compactness, and the Spectral Theory of
      Infinite Matrices}, available at
    http://www.reading.ac.uk/maths/research/maths-preprints.aspx

\bibitem[CW1]{CW1} X.\,Chen, Q.\,Wang, {\it Ideal structure of uniform
  Roe algebras of coarse spaces,} J.\ Func.\ Analys. {\bf 216}
(2004), 191--211. 

\bibitem[CW2]{CW2} X.\,Chen, Q.\,Wang, {\it Ghost ideal in uniform
    Roe algebras of coarse spaces,} Arch.\ Math. {\bf 84}
  (2005), 519--526

\bibitem[Cor]{Cor} H.\,O.\,Cordes, {\em Spectral theory of linear
    differential operators and comparison algebras }, Cambridge
  University Press, 1987.

\bibitem[DG1]{DG} M.\,Damak, V.\,Georgescu, {\it Self-adjoint
  operators affiliated to $C^*$-algebras,} Rev.\ Math.\ Phys. {\bf
  16} (2004), 257--280; this is part of 99--481 at
  http://www.ma.utexas.edu/mp\_arc/.

\bibitem[DG2]{DG2} M.\,Damak, V.\,Georgescu, {\it On the spectral
    analysis of many-body systems}, J.\ Func.\ Analys. (February
  2010); and preprint available at arXiv:0911.5126v1 at
  http://arXiv.org.

\bibitem[Dav]{D} E.\,B.\,Davies, {\it Decomposing the essential
    spectrum,} J.\ Func.\ Anal. {\bf 257} (2009) 506--536 and
  http://arxiv.org/abs/0809.5130.

\bibitem[Geo]{Ge} V.\,Georgescu, {\it On the spectral analysis of
    quantum field Hamiltonians}, J.\ Func.\ Analys. {\bf 245}
  (2007), 89--143; and preprint available at arXiv:math-ph/0604072v1
  at http://arXiv.org.

\bibitem[GG1]{GG1} V.\ Georgescu, S.\ Gol\'enia, {\em Isometries,
    Fock spaces, and spectral analysis of Schr\"odinger operators on
    trees}, J.\,Func.\,Analys. {\bf 227} (2005) 389--429; and
  preprint 04-182 at {http://www.ma.utexas.edu/mp\_arc/}.

\bibitem[GG2]{GG} V.\,Georgescu, S.\,Gol\'enia, {\it Decay
  preserving operators and stability of the essential spectrum },
  J.\,Op.\,Th. {\bf 59} (2008), 115--155; a more detailed
  version is http://arxiv.org/abs/math/0411489.

\bibitem[GI1]{GI0} V.\,Georgescu, A.\,Iftimovici, {\it Crossed
  products of $C^*$-algebras and spectral analysis of quantum
  Hamiltonians,} Comm.\,Math.\,Phys. {\bf 228} (2002), 519--560; see
  also 00-521 at http://www.ma.utexas.edu/mp\_arc/.

\bibitem[GI2]{GIc} { V.\,Georgescu, A.\,Iftimovici}, {\em
    $C^*$-algebras of quantum Hamiltonians}, in {Operator Algebras
    and Mathematical Physics}, eds.\,J.-M.\,Combes, J.\,Cuntz,
  G.A.\,Elliot, G.\,Nenciu, H.\,Siedentop, S.\,Stratila (Proceedings
  of the Conference Operator Algebras and Mathematical Physics,
  Constanta 2001), pp.\,123--167, Theta 2003; and preprint
  02-410 at {http://www.ma.utexas.edu/mp\_arc/}.

\bibitem[GI3]{GI} V.\,Georgescu, A.\,Iftimovici, {\it Localizations
  at infinity and essential spectrum of quantum Hamiltonians:
  I. General theory,} Rev.\,Math.\,Phys. {\bf 18} (2006),
  417--483; see also http://arxiv.org/abs/math-ph/0506051.

\bibitem[HP]{HP} U.\ Haagerup, A.\ Przybyszewska, {\it Proper
    metrics on locally compact groups, and proper affine isometric
    actions on Banach spaces}, preprint available at
  http://www.imada.sdu.dk/~haagerup/ (2006).

\bibitem[HM]{HM} { B.\,Helffer, A.\,Mohamed}, {\it Caract\'erisation
du spectre essentiel de l'op\'erateur de Schr\"odinger avec un champ
magn\'etique}, Ann.\,Inst.\,Fourier (Grenoble) {\bf 38}(2) (1988),
pp.\,95--112.

\bibitem[HLS]{HLS} N.\,Higson, V.\,Laforgue, G.\,Skandalis,
{\it Counterexamples to the Baum-Connes conjecture},
Geom.\,Funct.\,Anal. {\bf 12} (2002), 330--354.   

\bibitem[HPR]{HPR} N.\,Higson, E.K.\,Pedersen, J.\,Roe, {\it
    $C^*$-algebras and controlled topology}, K-Theory {\bf 11}
  (1997), 209--239.

\bibitem[LaS]{LS} { Y.\,Last, B.\,Simon}, {\it The essential
    spectrum of Schr\"odinger, Jacobi, and CMV operators},
  J. d'Analyse Math.  {\bf 98} (2006), 183-220; and preprint 05-112
  at {http://www.ma.utexas.edu/mp\_arc/}.

\bibitem[Mag]{Ma} A.\,Mageira, {\it $C^*$-alg\`ebres gradu\'ees par
    un semi-treillis}, th\`ese Universit\'e Paris 7, F\'evrier 2007;
  also available as  preprint number arXiv:0705.1961v1 at
  http://arxiv.org.

\bibitem[NY]{NY} P.\,Nowak, G.\,Yu, {\it What is Property A?},
  Notices of the AMS, {\bf 55} (2008), 474--475.

\bibitem[Pat]{Pa} A.\,T.\,Paterson, {\em Amenability\/}, Mathematical
  Surveys and Monographs {\bf 29}, AMS, Providence, Rhode Island
  1988.

\bibitem[Pi]{Pi} G.\,Pisier, {\it Similarity Problems and completely
    bounded maps} (second edition), Lect.\ Notes Math.\ {\bf 1618},
  Springer 2001.

\bibitem[RRR]{RRR} { V.\,S.\,Rabinovich, S.\,Roch, J.\,Roe},
{\it Fredholm indices of band-dominated operators},
Int.\,Eq.\,Op.\,Theory {\bf 49} (2004), 221-238.

\bibitem[RRS]{RRS} { V.\,S.\,Rabinovich, S.\,Roch, B.\,Silbermann},
  {\em Limit operators and their applications in operator theory},
Birkh\"auser, Series ``Operator Theory: Advances and Applications''
vol.\,150, 2004.

\bibitem[RS]{RS} H.\,Reiter, J.\,Stegman, {\it Classical harmonic
    analysis and locally compact groups}, Oxford Science
  Publications 2000.

\bibitem[Ro1]{R}  J.\,Roe, {\it Lectures on coarse geometry,} 
Am.\ Math.\ Soc. 2003.

\bibitem[Ro2]{Ro} {J.\,Roe},
{\it Band-dominated Fredholm operators on discrete groups},
{Int.\,Eq.\,Op.\,Theory} {\bf 51} (2005), 411-416.

\bibitem[STY]{STY} G.\,Skandalis, J.-L.\,Tu, G.\,Yu,
{\it The coarse Baum-Connes conjecture and groupoids},
Topology {\bf 41} (2002), 807--834.   

\bibitem[Tu]{Tu} J.-L.\,Tu, {\it Remarks on Yu's Property A for
    discrete metric spaces and groups,} Bull.\ Soc.\ Math.\ France
  {\bf 129} (2001), 115--139.

\bibitem[Wa]{Wa} Q.\,Wang, {\it Remarks on ghost projections and
    ideals in the Roe algebras of expander sequences,} Arch.\
  Math. {\bf 89} (2007), 459--465.

\bibitem[Wil]{Wi} D.P.\,Williams, {\it Crossed products of
  $C^*$-algebras \/}, Am.\ Math.\ Soc. 2007.

\bibitem[Yu]{Yu} G.\,Yu, {\it The coarse Baum-Connes conjecture for
    spaces which admit a uniform embedding into Hilbert spaces,}
  Inv.\ Math. {\bf 139} (2000), 201--240.

\end{thebibliography}
\end{document}